\documentclass{article}

\usepackage{PRIMEarxiv}

\usepackage[utf8]{inputenc} 
\usepackage[T1]{fontenc}    
\usepackage{hyperref}       
\usepackage{url}            
\usepackage{booktabs}       
\usepackage{amsfonts}       
\usepackage{nicefrac}       
\usepackage{microtype}      
\usepackage{fancyhdr}       
\usepackage{graphicx}       
\graphicspath{{media/}}     

\usepackage{hyperref}

\def\R#1{\texttt{#1}}

\usepackage{multirow}%
\usepackage{amsmath,amssymb}%
\usepackage{amsthm}%
\usepackage{mathrsfs}%
\usepackage[title]{appendix}%
\usepackage{xcolor}%
\usepackage{textcomp}%
\usepackage{float}
\usepackage{manyfoot}%
\usepackage{algorithm}%
\usepackage{algorithmicx}%
\usepackage{algpseudocode}%

\title{Mapping climate change awareness through spatial hierarchical clustering
\thanks{\textit{\underline{Citation}}: 
\textbf{Preprint version of "Mapping climate change awareness through spatial hierarchical clustering"}
} 
}

\author{
  Gianpaolo Zammarchi \\
  Department of Economics and Management  \\
  University of Cagliari, Italy \\
  \texttt{gianpaolo.zammarchi@unica.it}
   \And
  Paolo Maranzano \\
  Department of Economics, Management and Statistics (DEMS) \\ 
  University of Milano-Bicocca, Italy \\
  \texttt{paolo.maranzano@unimib.it}
}

\begin{document}
\maketitle

\begin{abstract}
Climate change is a critical issue that will be in the political agenda for the next decades. While it is important for this topic to be discussed at higher levels, it is also of paramount importance that the populations became aware of the problem. As different countries may face more or less severe repercussions, it is also useful to understand the degree of awareness of specific populations. In this paper, we present a geographically-informed hierarchical clustering analysis aimed at identify groups of countries with a similar level of climate change awareness. We employ a Ward-like clustering algorithm that combines information pertaining climate change awareness, socio-economic factors, climate-related characteristics of different countries, and the physical distances between countries. To choose suitable values for the clustering hyperparameters, we propose a customized algorithm that takes into account the within-clusters homogeneity, the between-clusters separation and that explicitly compares the geographically-informed and non-geographical partitioning. The results show that the geographically-informed clustering provides more stability of the partitions and leads to interpretable and geographically-compact aggregations compared to a clustering in which the geographical component is absent. In particular, we identify a clear contrast among Western countries, characterized by high and compact awareness, and Asian, African, and Middle Eastern countries having greater variability but still lower awareness.
\end{abstract}


\section{Introduction}
\subsection{Climate change and public awareness}
Climate change represents one of the greatest challenges that this and the future generations will face. Climate change poses serious risks to the planet's future and is a cause of environment's degradation. Historically, these changes have happened during a large part of Earth's history, but their evolution has always been rather slow if compared to the current trend. In many cases, the changes took hundreds if not millions of years to manifest themselves in a geologically detectable way \cite{loarie2009velocity}. Natural and anthropogenic emissions are both responsible of the today situations, but the scientific community agrees that humans have a key role in the extent and speed of these changes. Basically, the main drivers of these negative changes are associated to human behavior, especially when the rising global population became focused with the growth at all costs, which frequently leads to disregard the severe repercussions in natural systems and the consequences that follow from this \cite{loarie2009velocity}. 

Using fossil fuels for power, heating, and transportation has significantly raised greenhouse gas emissions and changed patterns of precipitations and temperature throughout the world. In 2022, the average worldwide temperature was approximately 0.86°C higher than the 20th century average and the last 50 years (almost) there was a continuous exceeding of the average values of previous years \cite{wang2023climate}. About weather patterns, and in particular precipitations, the effects of climate change are becoming more and more evident globally, as evidenced by the extreme weather events and related disasters that occur all over the world. Some of the most well-known events are the forest fires in Australia (like the devastating fires that occurred in Australia in 2019-2020 \cite{canadell2021multi}) and in the United States (especially in California \cite{goss2020climate}) or the intense rainfall in China \cite{sun2022understanding}, the droughts in South Africa \cite{meza2021drought}. These facts are now becoming common nowadays, but unfortunately they are not the only catastrophic events that are happening, just consider the melting of the glaciers and the reduction of snow on mountain tops, the rising sea levels, the changes to river flow patterns, and the risk of extinction of many species worldwide. In relation to this last point in fact, it has been estimated that, as worst case scenario, it is possible to witness a species loss between 16\% and 30\% \cite{roman2020recent}, which means millions of animals and plants to disappear \cite{IPBES2019}. Each one of these events will have several repercussions on the life of human beings and the animal species that live, hunt or base their life cycle on their environment.

As previously stated, all countries are impacted by climate change, however certain areas may be more exposed to particular effects. Given the seriousness of the problem, it is necessary to find solutions which, however, will probably require a long term to be implemented. For instance, reducing the amount of carbon dioxide in the atmosphere should be the primary goal to mitigate and carry out adaptation strategies for climate change. Other strategies include limiting development in floodplains, safeguarding naturally occurring wetlands and barrier islands, and fortifying vulnerable coastal communities with sea walls and levels \cite{titus1986greenhouse}. The goal of these policy tools is to make ecosystems and people more resilient to the fluctuations and changes in climate. Nevertheless, these initiatives may worsen disparities about the effects of climate change and provide challenges for regional policy coordination. For example, carbon intensive societies have a bigger absolute burden from attempts to reduce greenhouse gas emissions, since the costs associated with abatement, transition, and compliance are higher \cite{zahran2007ecological, edmonds2003costs}. In fact, as stated in the Global Climate Risk Index 2021 issued by the Germanwatch observatory, developing nations are less able to adapt, making them more vulnerable to the consequences of climate change \cite{GCRI2021}. In 2019, eight of the ten most severely impacted countries in terms of deaths and economic losses caused by extreme weather events (such as storms, floods, heatwaves, etc.) were, low- to lower-middle-income countries. Bahamas, Zimbabwe and Mozambique were three of the most impacted countries. Since countermeasures are not so easy and fast to implement, it is vital that governments of all countries take action to mitigate climate change. One of the possible pushes to act should come from the populations. For this to happen, it is necessary for individuals to become aware of the importance and urgency of interventions. Increasing public awareness on the causes and effects of climate change is of crucial importance because it can motivate policymakers to take action to cut greenhouse gas emissions and encourage individual behavioural adjustments. 

Given the different levels of vulnerability of countries around the world to the effects of climate change, the aim of this work was to examine how nations differ in terms of their degree of knowledge of the problem. Next, we present the findings of an analysis based on a spatially-constrained hierarchical clustering. To this aim, we compare different clustering scenarios in which combinations of socio-economic, climate, and geographic data were tested, while maintaining the settings of number of clusters and the mixing parameter $\alpha$ (for more information about $\alpha$ see Section \ref{hier_clust_geo_constr}).

\subsection{Related works on geographical clustering and our contribution}
\label{sec:related_workds}
With respect to the clustering methods, several solutions have been proposed in the literature to determine the optimal partition based on a homogeneity criteria based on dissimilarity (see e.g., \cite{gordon1996survey, ambroise1997clustering, liao2012clustering, miele2014spatially, pawitan2003constrained}).

In some cases, it makes sense to limit the range of feasible solutions, for instance by imposing contiguity constraints. Contiguity constraints can be both in time or in space and are the most prevalent kind. These constraints arise when an object in a cluster must be both similar to every other elements of a group and part of a continuous group of elements. Basically, if there is a path connecting each pair of elements in a cluster, then the cluster is said to be contiguous. Let \textbf{C} be the contiguity matrix where each entry $c_{ij}$ can be equal to 1 if the $i$-th element is contiguous to the $j$-th element and equal to 0 otherwise, then we can consider two clusters to be contiguous if the contiguity matrix shows a relationship ($c_{ij}$ = 1) between two elements (one from each cluster). However, many of the proposed methods work by considering that, if two elements are very similar but are located in different areas, the criterion based on spatial proximity will separate them into two different clusters. A possible solution is to consider soft constraints, so that the spatial contiguity does not create a sharp separation. In accordance with this logic, some methods proposed to merge the dissimilarity matrix derived from non-geographical features and the matrix of geographical distances \cite{oliver1989geostatistical, bourgault1992multivariate}. In this combination, the weights assigned to the geographical dissimilarities will provide more or less geographically contiguous clusters, but in this type of approach the issue is shifted to choosing the right weights.

In this paper, we apply a Ward-like hierarchical clustering technique with geographical or spatial constraints derived from the seminal work by \cite{chavent2018clustgeo}. This method employs a convex combination of two dissimilarity matrices, namely $D_0$ and $D_1$, which contain, respectively, information about a set of clustering features (in our case climate awareness, climate-related and socio-economic variables) and the geography of the area of interest. The two dissimilarities are linearly combined through a mixing hyperparameter which has to be selected carefully according to the field of application. The method works well with both Euclidean and non-Euclidean distances, which is the case when considering geographical distances. The general idea under the algorithm is to generate geographically compact clusters (i.e., with a marked spatial contiguity) of countries without excessively deteriorating the quality of the solution based on the set of the available features. Furthermore, as any other hierarchical clustering algorithm, the proposed method requires a proper identification of the optimal number of clusters to be considered. To deal with this, in the same spirit of \cite{chavent2018clustgeo}, we propose a practical procedure to identify suitable values of the mixing parameter and the number of clusters based and which takes into consideration several criteria, such as the within-clusters homogeneity, the between-clusters separation and explicitly comparing the geographically-informed and non-geographical results.

The remainder of the paper is organized as follows. In Section \ref{sec:material_methods} we illustrate the geographically-informed hierarchical clustering algorithm used in the empirical analysis and we present a new algorithm useful for the tuning of the clustering hyper-parameters and which extends previously existing procedures. In Section \ref{sec:results} we present the results of the clustering analysis using an extended dataset on climate change issues collected from various sources. In particular, we will present a main analysis based on the sole information provided by the climate change awareness survey 2022 and a set of robustness analyses taking into account the enriched database. Finally, in Section \ref{sec:conclusions} we provides some final remarks and define potential future developments.


\section{Material and methods}
\label{sec:material_methods}
In this section we illustrate how we collected the data for the carried out analyses. We describe in details the data sets used for the analyses and explain which methods are used to carry out these analysis, focusing in particular on hierarchical clustering.

\subsection{Data set}
\begin{figure}
	\centering
	\includegraphics[width=1\columnwidth]{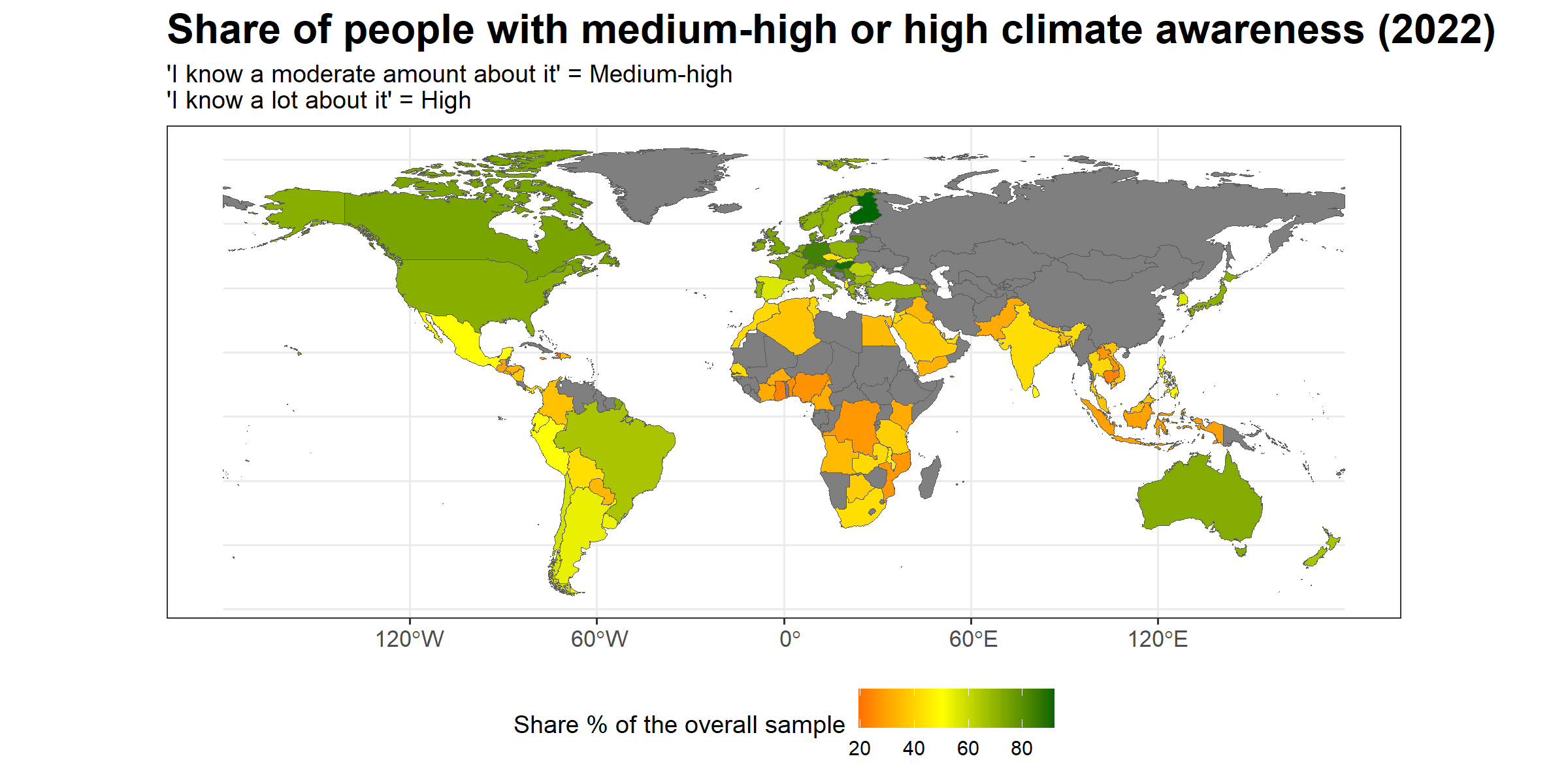}
	\caption{World map representing the share of people in the 2022 survey with medium-high and high climate change awareness at the country level. The share of people with medium and medium-high climate change awareness increases toward green and decreases toward red. The share was created by aggregating the participants declaring little or no awareness (“I have never heard of it", “I know a little about it"), while the complementary was created by aggregating the participants who declared moderate or high awareness of climate change (“I know a moderate amount about it", “I know a lot about it").}
	\label{fig:awareness_high}
\end{figure}

To quantify the level of awareness of each country we used the “International public opinion on climate change” survey. The two currently available editions refer to 2021 and 2022. The 2021 edition of the survey includes responses from 76,328 Facebook users located in 30 countries \cite{leiserowitz2021international}, while the 2022 edition includes responses from 108,946 Facebook users located in 95 countries worldwide \cite{leiserowitz2022international}. The survey was carried out in collaboration with “Data for Good at Meta" and the main aim was to examine people's knowledge, beliefs, attitudes, policy preferences, and behaviour related to climate change. We collected data pertaining to the “Climate Awareness” question which include five possible answers: “\textit{I have never heard of it}", “\textit{I know a little about it}", “\textit{I know a moderate amount about it}" and “\textit{I know a lot about it}", plus an option to decline to answer. To pursue our goals, we reclassified the previous statements into a binary response, namely the degree of climate change awareness, in which people answering “\textit{I have never heard of it}" and “\textit{I know a little about it}" were merged into \textit{Low or medium-low awareness about climate change}, whereas people answering “\textit{I have never heard of it}" and “\textit{I know a moderate amount about it}" were merged into \textit{I know a lot about it}. Then, for each country, we computed the share of respondents associated with the two reclassified answers. Figure \ref{fig:awareness_high} shows the world map based on the awareness levels of each of the 95 available countries. The map shows that the US, Canada, Brazil, Australia, New Zealand, and almost all Europe have high awareness values, while African countries and Asian countries have rather low values of awareness (with some exceptions, e.g. Japan).

In addition to the survey data, we enriched the dataset with a set of climate-related and socio-economic variables collected from the Climate Change Dashboard of the International Monetary Fund, the Climate Change Knowledge Portal, the Penn World Table, the Global Carbon Atlas, and the Our World in Data web portal. For every country and variable we considered the last available data in the period 2021-2022. In Table \ref{tab:variables} we present the complete list of variables used in the analysis, the source from which they were obtained, a brief explanation of the meaning of the variable, and their classification into socio-economic or climate-related variables.

\begin{table}[!ht]
\caption{Variables included in the analysis and their sources}\label{tab:variables}
\begin{tabular}{lll}
\toprule
\textbf{Variable name} & \textbf{Variable description} & \textbf{Source} \\
\midrule
\multirow{2}{*}{CarbonIntens\_Electr} & Grams of carbon dioxide-equivalents emitted & \multirow{2}{*}{Our World In Data}\\
& per kWh of electricity generated &  \\
\multirow{2}{*}{EnerIntens\_PrimEnergy} & Amount of energy needed to   produce one unit & \multirow{2}{*}{Our World In Data}\\
& of economic output (megajoules per dollar) &  \\
HDI & Human Development Index & Our World In Data \\
csh\_g & Share of government consumption at current PPPs & Penn World Table \\
\multirow{2}{*}{rgdpna} & Real GDP at constant 2017 national prices & \multirow{2}{*}{Penn World Table} \\
& (in mil. 2017US\$) &  \\
EmpRate & Employment rate & Penn World Table \\
\multirow{2}{*}{TerritorialEmiss\_IntensGDP\_KgThs} & Ratio between carbon dioxide emissions and & \multirow{2}{*}{Global Carbon Atlas} \\
& real GDP (2017) &  \\
\multirow{2}{*}{TradeOpenness} & Sum of merchandise imported and & \multirow{2}{*}{Penn World Table}\\   
& exported at current PPPs (\%) &  \\
\hline
cdd & Maximum number of consecutive dry days & CCKP \\
hd30 & Number of Hot Days (Tmax \textgreater 30°C) & CCKP \\
pr & Precipitation & CCKP \\
tx84rr & Excess Mortality & CCKP \\
wsdi & Warm Spell Duration Index & CCKP \\
tas & Average Mean Surface Air   Temperature & CCKP \\
\hline
\multirow{2}{*}{Medium-low and low climate awareness} & Share of people with no or little awareness & Climate Change \\
& about climate change & Opinion Survey (2022) \\
\multirow{2}{*}{Medium-high and high climate awareness} & Share of people with moderate or high awareness & Climate Change \\
& about climate change & Opinion Survey (2022)\\
\bottomrule
\multicolumn{2}{l}{\small Abbreviations: CCKP, Climate Change Knowledge Portal; pop., population} & \\
\end{tabular}
\end{table}

\subsection{Ward's hierarchical clustering} \label{sec:clustering}
The first goal we set is to investigate variations in the patterns of climate change awareness across time. To this aim, we incorporated information from the 30 countries of the 2021 edition and matched with the 2022 edition, in order to include only countries that were in both survey editions. Using the four response items, we constructed a distance matrix in order to group countries based on their awareness levels. As shown in Equation \ref{eq:ward}, single observations were combined into clusters using the Ward's approach.
\begin{equation}
\begin{aligned}
\frac{\left|A\right|\cdot\left|B\right|}{\left|A\cup B\right|}\left|\left|\mu_A - \mu_B\right|\right|^2 = 
\sum_{x\in A\cup B}\left|\left|x-\mu_{A\cup B}\right|\right|^2-\sum_{x\in A}\left|\left|x-\mu_A\right|\right|^2 - \sum_{x\in B}\left|\left|x-\mu_B\right|\right|^2  
\end{aligned}
\label{eq:ward}
\end{equation}
where A and B are the two sets of observations. We created a hierarchical cluster using the \R{hclust} function in the \R{stats} \R{R} software \cite{CRAN}. The \R{dendextend} R package \cite{dendextend} will be used to compare the two hierarchical clusters in order to highlight cluster variations occurred between the 2021 and the 2022 editions. Let $E$ be the \textit{entanglement} value between the lines connecting the countries in the two hierarchical plots, and $v_1 = [1, 2, \dots, n]$ is the vector corresponding to the order from $1$ to $n$ (where $n$ is the number of elements to be clustered) in which the labels/elements names appear on the left side of the plot. Finally, $v_2$ is the vector containing the same elements as $v_1$, required to carry out the comparison. Moreover, let $|| \ . \ ||_L$ be the L-norm distance between these two vectors. To normalize the value obtained it is necessary to compute a maximum value $M$ for this distance, which is the value that refers to the worst possible scenario when one set of labels is the opposite of the other. Formally, the \textit{entanglement} can be written as in Equation \ref{eq:entanglement}:
\begin{equation}
E = \frac{\left|\left|v_1-v_2\right|\right|_L}{M} \quad \text{with} \ E \in [0, 1]
\label{eq:entanglement}
\end{equation}
The value obtained is useful to get an idea of how similar the two classifications are. In fact, the two extreme scenarios are when $E = 0$, which indicates a perfect match between the two sets of labels or $E = 1$, which is the case when the right set of labels is the exact opposite of the left side. Since the 2021 edition of the awareness survey only included 30 countries, we decided that the following parts of the analysis would be performed only on the 2022 data in order to have a larger number of countries to analyse.

\subsection{Hierarchical clustering with geographical constraints}
\label{hier_clust_geo_constr}
The original survey data were then augmented by a set of climate-related and socio-economic features at the country level collected from the sources listed in Section \ref{sec:material_methods} and summarized in Table \ref{tab:variables}. The expanded data set is used as database to carry out a hierarchical clustering with geographic constraints to explore the climate change awareness patterns. To do so, we considered the spatial hierarchical clustering algorithm proposed by \cite{chavent2018clustgeo}, which relies on a linear combination of the dissimilarity in the feature space and the geographical dissimilarity (i.e., distance in the spatial coordinates) to obtain the partitioning of the units under proximity constraint. Specifically, this method makes use of two dissimilarity matrices, that is, a feature-based distance matrix $D_0 = [d_{0,ij} ]_{i,j=1,\dots,n}$ and a geographical distances matrix $D_1 = [d_{1,ij} ]_{i,j=1,\dots,n}$. The interpretation of the two matrices is straightforward: while the $D_0$ matrix  provides information on dissimilarity in the “feature space”, the $D_1$ distance matrix embeds information in the geographical “constraint space”. Notice that to compute the geographical distance we used geodetic distance function, which consider the shortest path between two points on a curved surface, such as an ellipsoid or sphere, which in this case is the Earth's surface. Also, as we will see later, the feature space in our application will comprehend a set of climate change awareness, socio-economic, and climate-related variables at the national level for 2022.

The clustering is performed following a Ward-like hierarchical strategy in which the dissimilarity matrix is given by a convex linear combination of $D_0$ and $D_1$, linearly related by a mixing parameter $\alpha$, which controls how much importance is given to each one of the two matrices in the clustering procedure. When the mixing parameter $\alpha$ approaches 1, contiguous units will be forced to cluster together as the weight of the features dissimilarity becomes negligible and the only relevant dimension is the geographical distance. On the contrary, for values of $\alpha$ approaching 0, the spatial constraint will become progressively weaker and the feature-based dissimilarities will prevail.

Let $I_\alpha$ denote the mixed pseudo inertia, then the mixed pseudo inertia of cluster $C_k^\alpha$ can be defined as \ref{eq:inertia_ck} 
\begin{equation}
\begin{aligned}
I_\alpha\left(C_k^\alpha\right) = \left(1-\alpha\right)\sum_{i\in C_k^\alpha}\sum_{j\in C_k^\alpha}{\frac{w_iw_j}{2\mu_k^\alpha}d_{0,ij}^2} \quad + 
\alpha\sum_{i\in C_k^\alpha}\sum_{j\in C_k^\alpha}{\frac{w_iw_j}{2\mu_k^\alpha}d_{1,ij}^2}
\end{aligned}
\label{eq:inertia_ck}
\end{equation}
where $w_i$ is the weight of the $i$-th observation for $i = 1,..., n$, $\mu_k^\alpha\ =\sum_{i\in C_k^\alpha} w_i$ is the weight of $C_k^\alpha, d_{0,ij}^2$ is the normalized $n\ \times n$ dissimilarity matrix between observations $i$ and $j$ in $D_0$, and $d_{1,ij}^2$ is the analogous in $D_1$.

As in the classical Ward-like hierarchical procedure, the scope here is to minimize the convex combination of the homogeneity criterion calculated with $D_0$ and the homogeneity criterion calculated with $D_1$. To this extent, recall that when $\alpha$ increases, the homogeneity computed with $D_0$ decreases while the homogeneity computed with $D_1$ increases. Also, recall that the degree of homogeneity associated with a partition is measured through its pseudo inertia; in the mixed setup by \cite{chavent2018clustgeo}, as the mixed pseudo inertia of cluster $C_k^\alpha$ decreases, the units belonging to cluster $C_k^\alpha$ are more homogeneous (i.e., $\downarrow I_\alpha\left(C_k^\alpha\right)$ then $\uparrow$ homogeneity within $C_k^\alpha$). When considering a partitioning into $K$ clusters, the overall homogeneity is given by the mixed within-cluster pseudo inertia, namely $W(\mathcal{P}_K)$, which is computed as
\begin{equation}
W_\alpha(\mathcal{P}_K^{\alpha})=\sum_{k=1}^{K} I_\alpha(\mathcal{C}_k^{\alpha})
\end{equation}
In this framework, in order to obtain a high degree of homogeneity of the partition $\mathcal{P}_K$, the clusters are formed by minimizing the mixed within-cluster pseudo inertia.
 
\subsection{Tuning of the hyperparameters $\alpha$ and $K$}
In the typical context of hierarchical clustering a crucial issue is how to determine the optimal number of clusters, that is, $K^*$. Several criteria are available in the literature, and ranging from graphical methods, such as the elbow method, to analytical tools, such as distance-based indicators (e.g., the Calinski–Harabasz index or the silhouette width as in \cite{AkhanliHennig2020}) or model-based indicators based on the likelihood function (e.g., likelihood criteria as in \cite{RossbroichEtAl2022}). In our framework, the complexity is even higher as in addition to the number of clusters we have to choose a suitable value for the mixing parameter $\alpha$, that is, $\alpha^*$. Also, we recall that it exists a mutual relationship among $\alpha$ and $K$ as the mixing parameter depends on the number of clusters, but also the mixing parameter can influence the number of clusters.

Among the available ways to define $\alpha^*$, we recall the proposal by \cite{chavent2018clustgeo} and \cite{MorelliEtAl2024}, both based on a sequential approach in which $\alpha$ is determined by conditioning on the number of clusters $K$. The former proposal defines $\alpha$ that best compromises between loss of feature-based and loss of geographical homogeneity. In other words, \cite{chavent2018clustgeo} determine $\alpha$ such that it increases the spatial contiguity without deteriorating too much the quality of the solution based on the feature space. Such objective is obtained by minimizing the distance among the proportion of the total mixed pseudo inertia explained by the partition $\mathcal{P}_K$ in $K$ clusters normalized with respect to the $D_0$ matrix (that is, compared to the case of clustering based only on the non-spatial features), namely $\tilde{Q}_{D_0}(\mathcal{P}_K^{\alpha})$, and the proportion normalized with respect to the $D_1$ matrix (that is, compared to the case of clustering based only on the geographical distance), namely $\tilde{Q}_{D_1}(\mathcal{P}_K^{\alpha})$. Conversely, \cite{MorelliEtAl2024} propose to select the mixing parameter $\alpha$ which jointly maximizes the amount of pseudo inertia explained from both the socio-economic features and the geographical information, weighted by the cumulated spatial and socio-economic pseudo inertia embedded the data. A comprehensive discussion about the computation and the interpretation of the two methods, as well as their relationship with the clustering-related metrics, is provided in Section 4 of \cite{MorelliEtAl2024} and Section 3 of \cite{chavent2018clustgeo}.

In \cite{MorelliEtAl2024}, the authors propose several sequential algorithms to show how the selection of both $K$ and $\alpha$ can be performed; in particular, conditioning on an initial value of $K$, the mixing parameter $\alpha^*$ is find according to the criterion above, and then the number of clusters $K^*$ is re-optimized using one or more clustering criteria defined by the user. Here, we propose a procedure that combines the \cite{MorelliEtAl2024} and \cite{chavent2018clustgeo} criteria and extends the set of clustering indicators used to select a suitable number of clusters and the mixing parameters. The proposed procedure is described in Algorithm \ref{algorithm:sp}.

\begin{algorithm}
\caption{Hierarchical Spatial Clustering: choice of the hyperparameters $\alpha$ and $K$}\label{alg:cap}
\begin{algorithmic}
\vspace{0.2cm}
\State \underline{\textbf{Step 1: define the algorithm's inputs}}
\State Define as $D_0 = [d_{0,ij} ]_{i,j=1,\dots,n}$ the feature dissimilarity matrix 
\State Define as $D_1 = [d_{1,ij} ]_{i,j=1,\dots,n}$ the spatial dissimilarity matrix  
\State Define as $K_{max}$ the maximum number of clusters
\State Define as $\underline{\alpha} \in [0,1]$ a sequence of mixing parameters
\vspace{0.2cm}
\State \underline{\textbf{Step 2: identify the set of potential optimal mixing parameters}}
\For {$K=2,\dots, K_{max}$}
    \For{$\alpha \in \underline{\alpha}$}
        \State Compute the linear combination of the two dissimilarity matrices $D(\alpha)=(1-\alpha)D_0+\alpha D_1$;
        \State Compute the $\mathcal{P}_K^{\alpha}=$ partition in $K$ clusters according to Ward's hierarchical algorithm on $D$;
        \State Compute the weighted average of the explained mixed pseudo inertia $\bar{Q}(\mathcal{P}_K^{\alpha})$ as in \cite{MorelliEtAl2024} and the two normalized pseudo inertia $\tilde{Q}_{D_0}(\mathcal{P}_K^{\alpha})$ and $\tilde{Q}_{D_1}(\mathcal{P}_K^{\alpha})$ as in \cite{chavent2018clustgeo};
    \EndFor
    \State Select the best $\alpha$ for each $K$ such that $\alpha_{K,max}^*=argmax_{\alpha} \bar{Q}(\mathcal{P}_K^{\alpha})$ (that is, according to the criterion by \cite{MorelliEtAl2024}) and the best $\alpha$ for each $K$ such that $\alpha_{K,min}^*=argmin_{\alpha}|\tilde{Q}_{D_0}(\mathcal{P}_K^{\alpha}) - \tilde{Q}_{D_1}(\mathcal{P}_K^{\alpha})|$ (that is, according to the criterion by \cite{chavent2018clustgeo})
\EndFor
\vspace{0.2cm}
\State \underline{\textbf{Step 3: define the optimal combination of number of clusters and mixing parameters}}
\State Choose $K^*$ (evaluated at the corresponding $\alpha_{K,max}^*$ and $\alpha_{K,min}^*$) based on a set of hierarchical clustering criteria, such as the Silhouette index, the Dunn's index, the C-index, the Calinski-Harabasz's index, and the McClain-Rao's index. In particular, consider the combination of three potential rule-of-thumb: (1) graphical analysis of the indices for varying $\alpha$'s and $K$'s; (2) majority voting rule on the absolute values of the clustering criteria; (3) majority voting rule on the gain/loss of the indices evaluated at $\alpha_{K,max}^*$ and $\alpha_{K,min}^*$ (i.e., geographically-informed clustering) compared to the case of the indices at $\alpha = 0$ (i.e., non-spatial clustering).
\vspace{0.2cm}
\end{algorithmic}
\label{algorithm:sp}
\end{algorithm}

The algorithm works in a three-stage setting. In the first stage, the user has to define the hierarchical clustering's inputs (i.e., the feature dissimilarity and spatial dissimilarity matrices, the maximum number of clusters to be considered, and a sequence of candidate mixing parameters to weight the matrices. In the second step we identify the set of potential optimal mixing parameters based on both the \cite{MorelliEtAl2024} (denoted as $\alpha_{K,max}^*$) and \cite{chavent2018clustgeo} (denoted as $\alpha_{K,min}^*$) criteria for all the considered $K$'s. In the third step we use a combination of three potential rule-of-thumb to identify the definitive number of clusters $K^*$, that is, graphical analysis of the indices for varying $\alpha$'s and $K$'s; majority voting rule \cite{NbClust2014} on the absolute values of a set of clustering criteria (i.e., Silhouette index \cite{Rousseeuw1987}, the Dunn's index \cite{Dunn1974}, the C-index \cite{HubertLevin1976}, the Calinski-Harabasz's index \cite{CH1974}, and the McClain-Rao's index \cite{McClainRao1975}); majority voting rule on the gain/loss of the previous set of indices evaluated at $\alpha_{K,max}^*$ and $\alpha_{K,min}^*$ (i.e., geographically-informed clustering) compared to the case in which we consider only the feature space and ignore the geographical information (i.e., $\alpha = 0$). Specifically, the latter rule-of-thumb is based on the computation of the percentage gain or loss obtained by using a geographically-informed clustering approach instead of the geography-free clustering, that is when fixing $\alpha = 0$. For a given index, the gain/loss is obtained as the following percentage variation:
\begin{equation}
GL = \frac{Index_{\alpha^*_K} - Index_{\alpha_K = 0}}{Index_{\alpha_K = 0}} \times 100.
\end{equation}
Notice that the definition of gain or loss depends on how each index is optimized. Indeed, when considering the C-index and the McClain-Rao index, the optimal number of clusters is the minimizer, while for Silhouhette, Calinski-Harabasz, and Dunn indices the optimal value is $K$ such that the indicators are maximized. Thus, for the former two a gain (loss) is detected when the indicator associated with the geographically-informed clustering returns a lower (greater) value compared to the non-spatial case. Conversely, for the latter three a gain (loss) is detected when the indicator associated with the geographically-informed clustering returns a greater (lower) value compared to the non-spatial case. In other words, for Silhouhette, Calinski-Harabasz, and Dunn the spatial information induces improvements for positive variations, while for the C-index and the McClain-Rao index an improvement is obtained when negative variations occur. We also remark that by expressing the improvements with respect to the geography-free case, in the unlucky situation of losses, we can both quantify the amount of such losses and eventually choose a set of hyperparameters that guarantees performances similar to the non-spatial clustering.

To sum up, the optimal combination ($\alpha^*_K$,$K^*$) is selected such that it simultaneously takes into account the within-clusters homogeneity (both when selecting the mixing parameter using the pseudo inertia and when computing the clustering criteria), between-clusters separation (e.g., the Calinski-Harabasz's index consider both within-cluster and between cluster variability \cite{AkhanliHennig2020}), and an explicit comparison between the geographically-informed and non-geographical clusterings (when computing the gain/loss of the indicators).


\section{Results} \label{sec:results}
\subsection{A preliminary comparison of the 2021 and 2022 awareness surveys}
\begin{figure}[!ht]
	\centering
	\includegraphics[width=1\columnwidth]{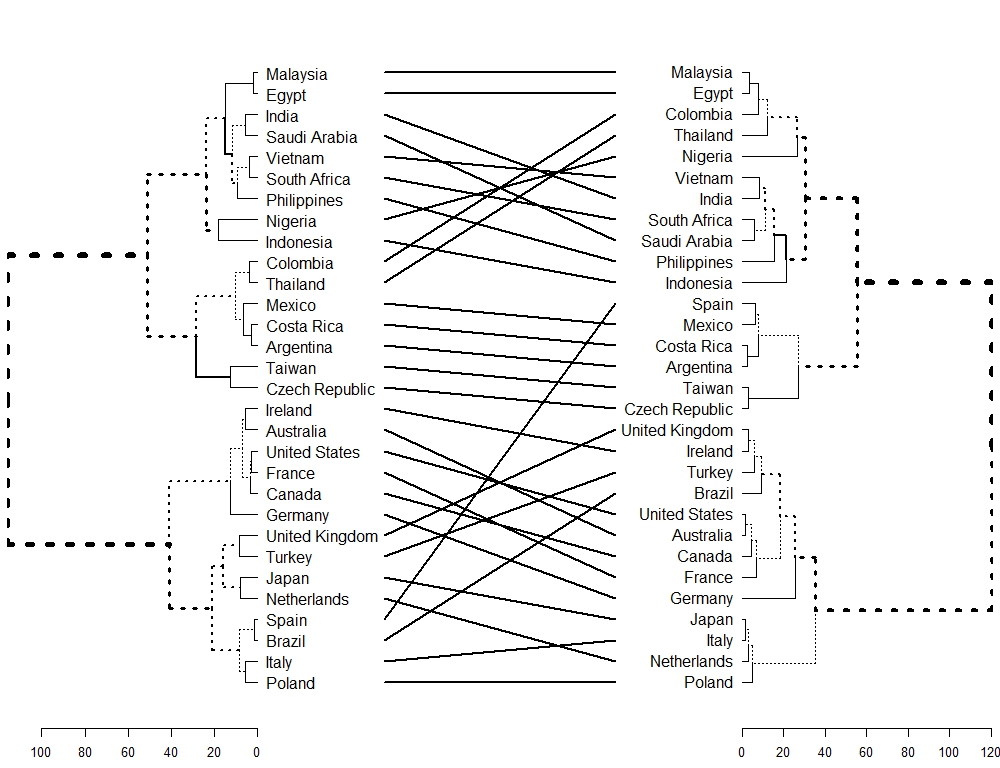}
	\caption{Comparison between awareness measured in 2021 and 2022}
	\label{fig:comparison21_22}
\end{figure}

The first part of the analysis aimed at comparing the two hierarchical clustering plots produced for 2021 and 2022 to evaluate whether there was a variation in the countries that were part of the clusters (Figure \ref{fig:comparison21_22}). Based on the elbow method, an optimal $k = 3$ number of clusters was identified for both data sets. The comparison of the two hierarchical clusters can be numerically evaluated using the method described in Section \ref{sec:clustering}. An entanglement value $E$ = 0.29 was obtained from this analysis, suggesting a modest variation in cluster membership between the two years. Among the countries showing different levels of awareness, Thailand and Colombia relocated from Cluster 2 to Cluster 3 (i.e. in a cluster with poorer awareness of climate change) and Spain relocated from Cluster 1 to Cluster 2, which is a cluster with a lower awareness level.

\subsection{Hierarchical spatial clustering using only climate change awareness information}
\begin{figure}
	\centering
	\includegraphics[width=1\columnwidth]{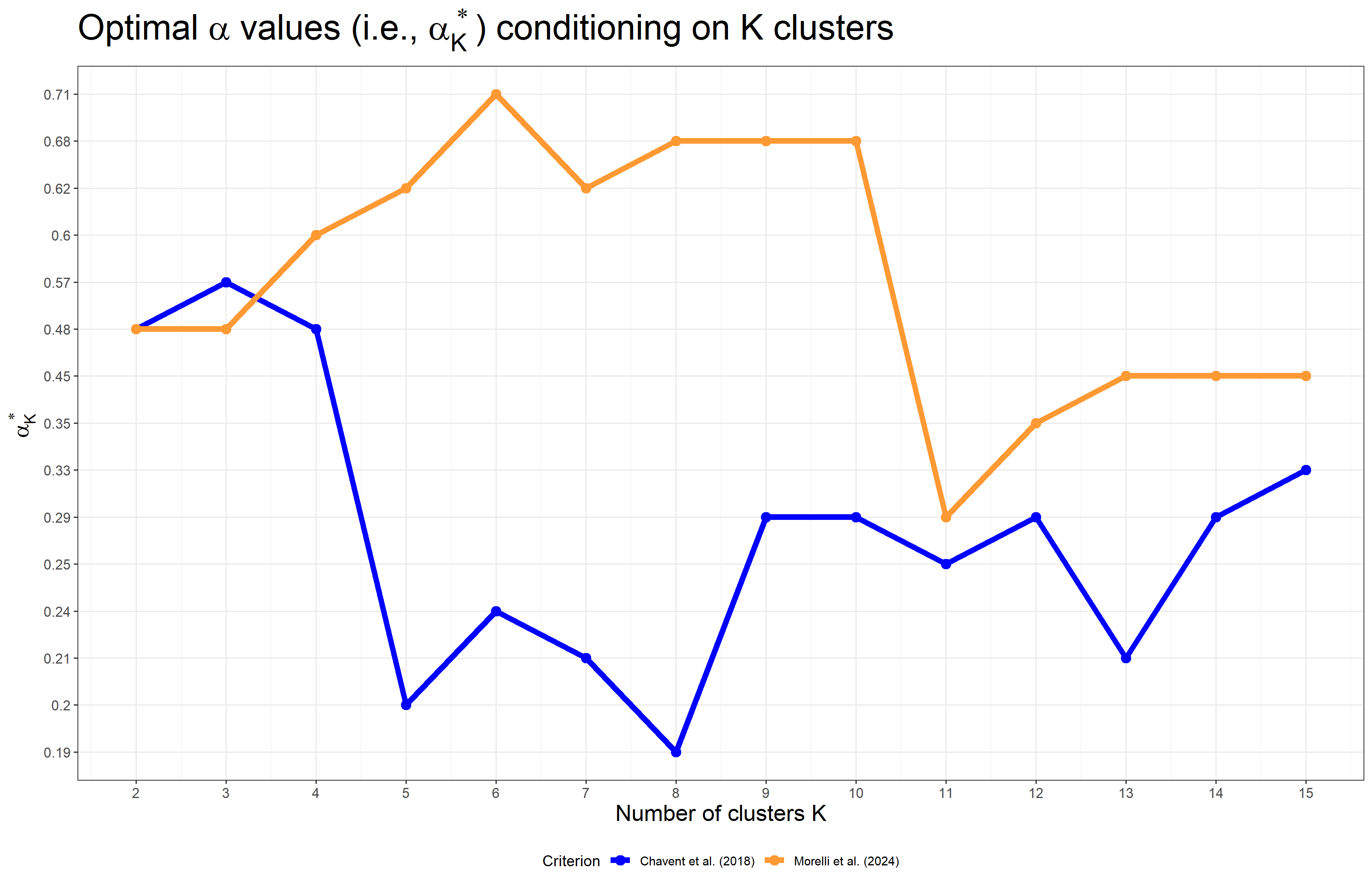}
	\caption{Optimal values of the mixing parameter $\alpha$ conditioning on $K$ clusters (that is, $\alpha_K^*$) obtained using the criterion by \cite{MorelliEtAl2024} ($\alpha_{K,max}^*$ in orange) and by \cite{chavent2018clustgeo} ($\alpha_{K,min}^*$ in blue).}
	\label{fig:optim_alpha}
\end{figure}

In the second part of the analysis we focused on the awareness survey conducted in 2022, gathering data from 95 areas worldwide, thus allowing to present a much broader view of the world situation. In Figure \ref{fig:optim_alpha} we show the optimal $\alpha$ values (i.e. $\alpha_K^*$) computed for each value of $K$ from 2 to 15. The $(K_{max}-1)$ $\alpha_K^*$ values were obtained using the criteria proposed by \cite{MorelliEtAl2024} (orange) and \cite{chavent2018clustgeo} (blue). From the figure we can easily infer that while the \cite{chavent2018clustgeo}'s criterion assigns a low weight to the geographical information (i.e., $\alpha_{K,min}^*$ always lies between 0.19 and 0.48, with higher values for small amount of clusters), the criterion by \cite{MorelliEtAl2024} suggest a more relevant role for the spatial constraint.

\begin{figure}
	\centering
	\includegraphics[width=1\columnwidth]{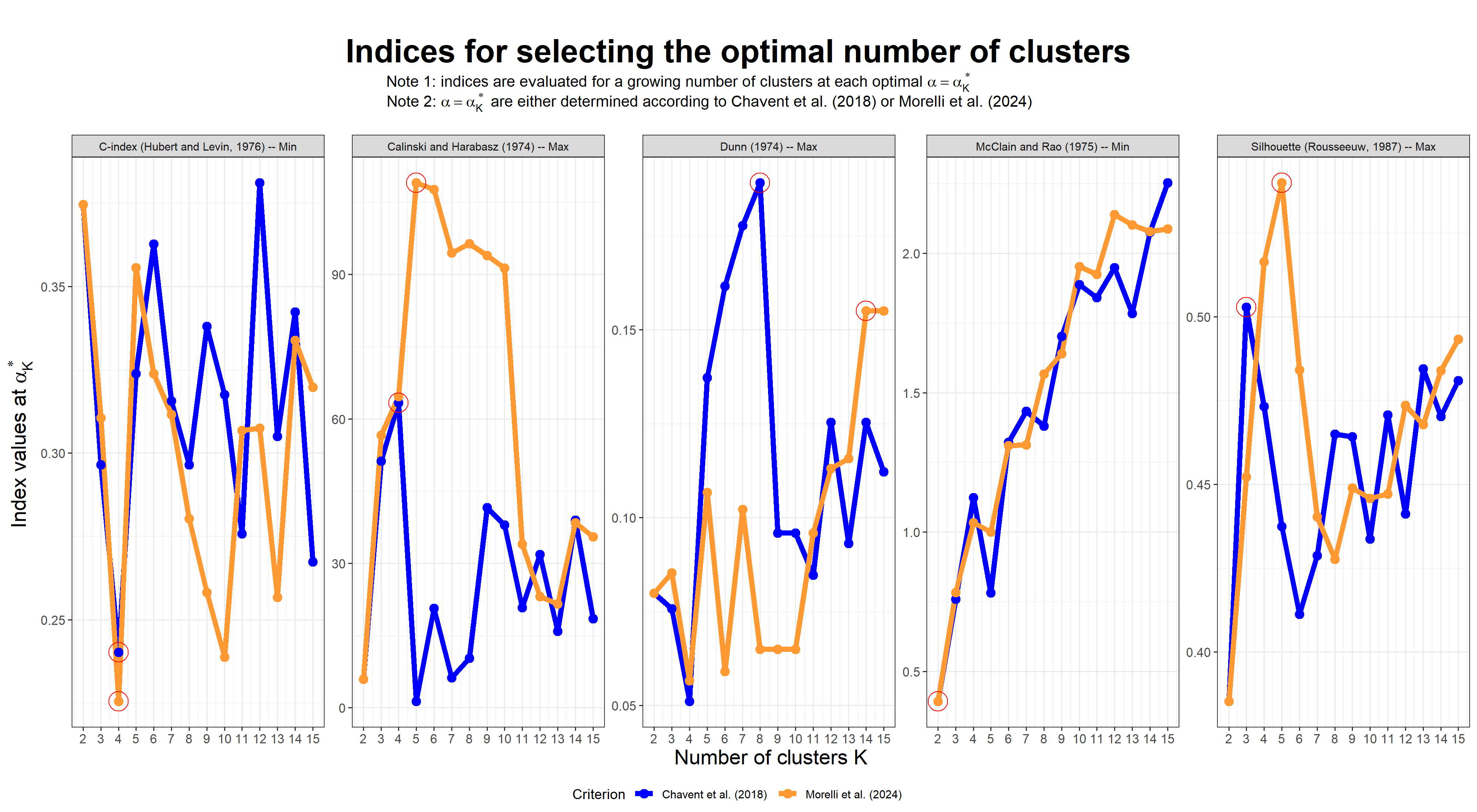}
	\caption{Estimates of the Silhouette index, the Dunn's index, the C-index, the Calinski-Harabasz's index, and the McClain-Rao's index for $K$ clusters evaluated at the corresponding $\alpha_{K,max}^*$ (orange) and $\alpha_{K,min}^*$ (blue). Red circles identifies the optimal values, that is, the minimizer for C-index and McClain-Rao's indices and the maximizer for Silhouette, Dunn's and Calinski-Harabasz's indices.}
	\label{fig:indices_optimK}
\end{figure}

As shown in Figure \ref{fig:indices_optimK}, for a given value of $K$ and $\alpha_K^*$ it is possible to obtain a specific index value for each of the five proposed metrics. These metrics were the Silhouette index, the Dunn's index, the C-index, the Calinski-Harabasz's index, and the McClain-Rao's index. However, not all the metrics work in the same manner. In fact, for the C-index and the McClain-Rao index the optimal value is the one corresponding to the minimizer, while for Silhouette, Dunn and Calinski-Harabasz indices indicate that the optimal value is the maximum. Taking these facts into account, Figure \ref{fig:indices_optimK} can be used to establish which value of $K$ corresponds to the optimal value, that we will indicate as $K^*$. Potentially, any combination of metrics (i.e., the indices) and methods (i.e., \cite{chavent2018clustgeo} and \cite{MorelliEtAl2024}) could return a different value for $K^*$, thus leading to a maximum of ten optimal $K^*$. In our analysis only six different values were obtained. These values were 2 (two times), 3 (one time), 4 (three times), 5 (two times), 8 (one time), and 14 (one time).

\begin{figure}
	\centering
	\includegraphics[width=1\columnwidth]{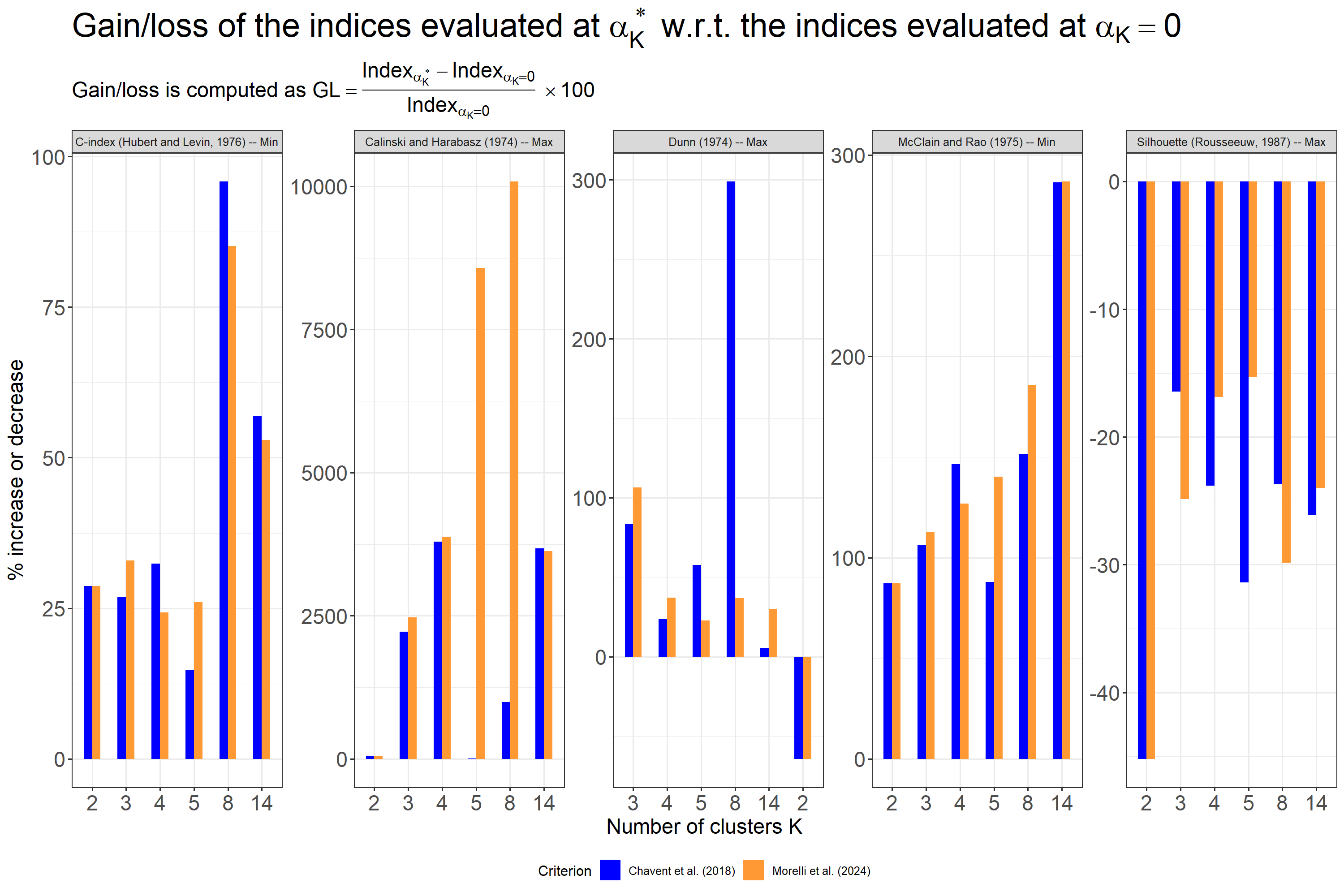}
	\caption{Estimated percentage gain (positive values) or loss (negative values) of the Silhouette, Dunn's, C-index, Calinski-Harabasz's index, and McClain-Rao's indices evaluated at the corresponding $\alpha_{K,max}^*$ (orange) and $\alpha_{K,min}^*$ (blue) with respect to the indices evaluated at $\alpha = 0$. Percentage gain/loss is computed as $GL = \frac{Index_{\alpha^*_K} - Index_{\alpha_K = 0}}{Index_{\alpha_K = 0}} \times 100$; thus, positive values have to be read as the percentage increase of the indices w.r.t. the baseline case of index at $\alpha = 0$, while negative values have to be read as the percentage decrease of the indices w.r.t. the baseline case of index at $\alpha = 0$.}
	\label{fig:gain_loss_alpha0}
\end{figure}

In Figure \ref{fig:gain_loss_alpha0} we show the estimated percentage gain or loss of the five metrics when using the geographical constraint (i.e., indices are evaluated at $\alpha=\alpha_{K,min}^*$ or $\alpha=\alpha_{K,max}^*$) compared to the case of geography-free clustering (i.e., indices are evaluated at $\alpha=0$). For all the indices except the Silhouette, we observe that the optimized $\alpha_K^*$ produce larger values in the absolute scale, leading to positive percentage increments. The magnitude of the gain varies based on the metrics and the specific $K$ values. The Silhouette index is the only on showing negative variations for all the considered combinations of $K$ and $\alpha$. In many cases, the cases $K=4$ and $K=5$ lead to marked gains or moderate losses. For instance, regarding the Calinski-Harabasz's index, using the $\alpha_K^*$ values computed using the method proposed by \cite{MorelliEtAl2024} allows to obtain a particularly high gain for $K = 5$ and $K = 8$, while moderate improvements are detected when considering $K=4$ regardless the selected method. Also, $K=4$ and $K=5$ evaluated at $\alpha_{K,max}^*$ provide the minor losses for C-index and Silhouette index. Considering the partial concordance among the optimal solution, we believe that a good compromise can be achieved by choosing $K^*=4$ and $K^*=5$ evaluated using the methods by \cite{MorelliEtAl2024}. Below we present the results related to these two cases when considering the geographical constraint and when ignoring the spatial dimension. 

\begin{figure}
	\centering
	\includegraphics[width=1\columnwidth]{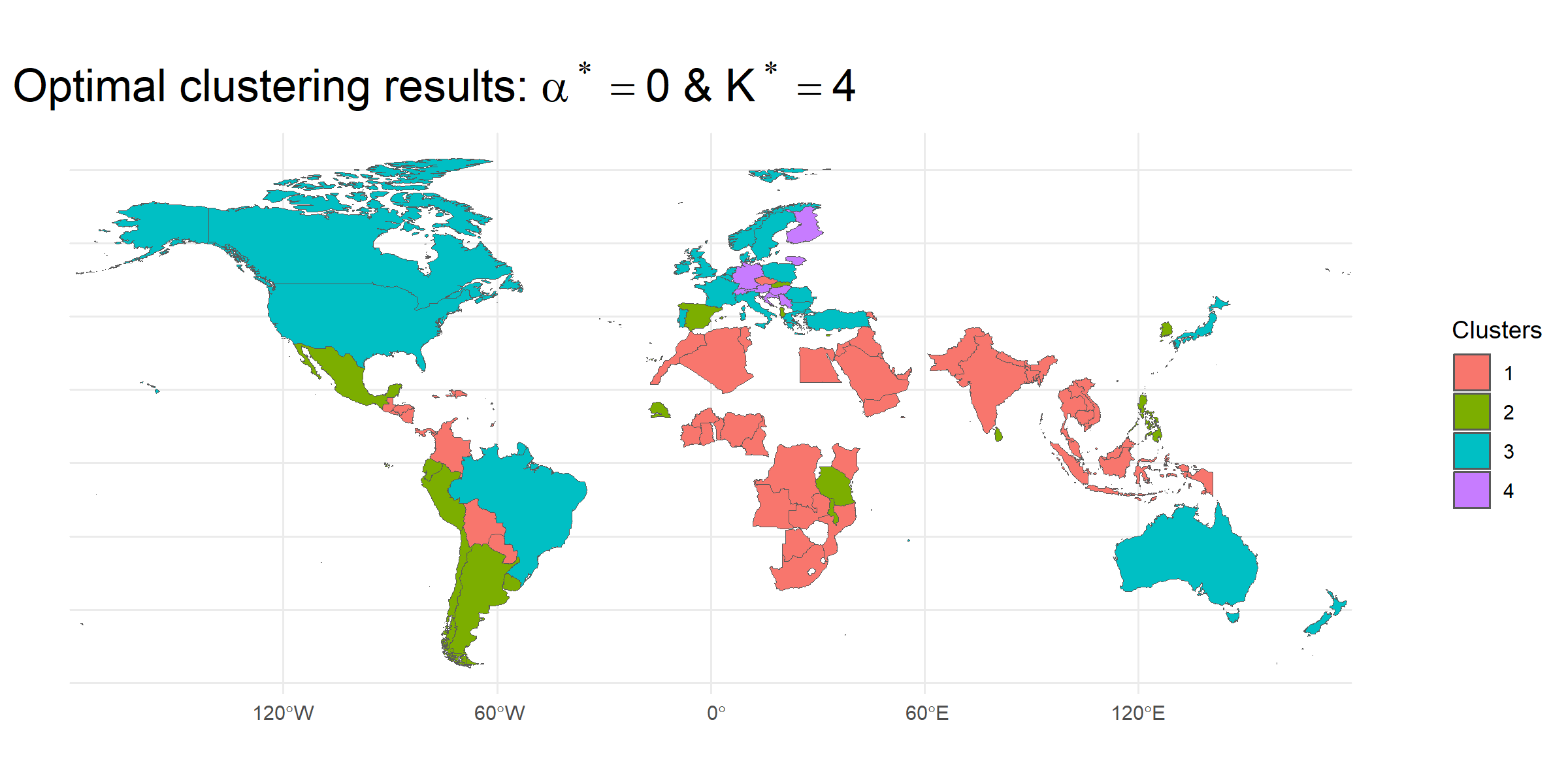}
	\caption{Map of clustering partitions obtained by setting $\alpha = 0$ and $K = 4$ and using as data the share of people in the 2022 survey with medium-low and low climate change awareness.}
	\label{fig:map_k4_a0}
\end{figure}

\begin{figure}
	\centering
	\includegraphics[width=1\columnwidth]{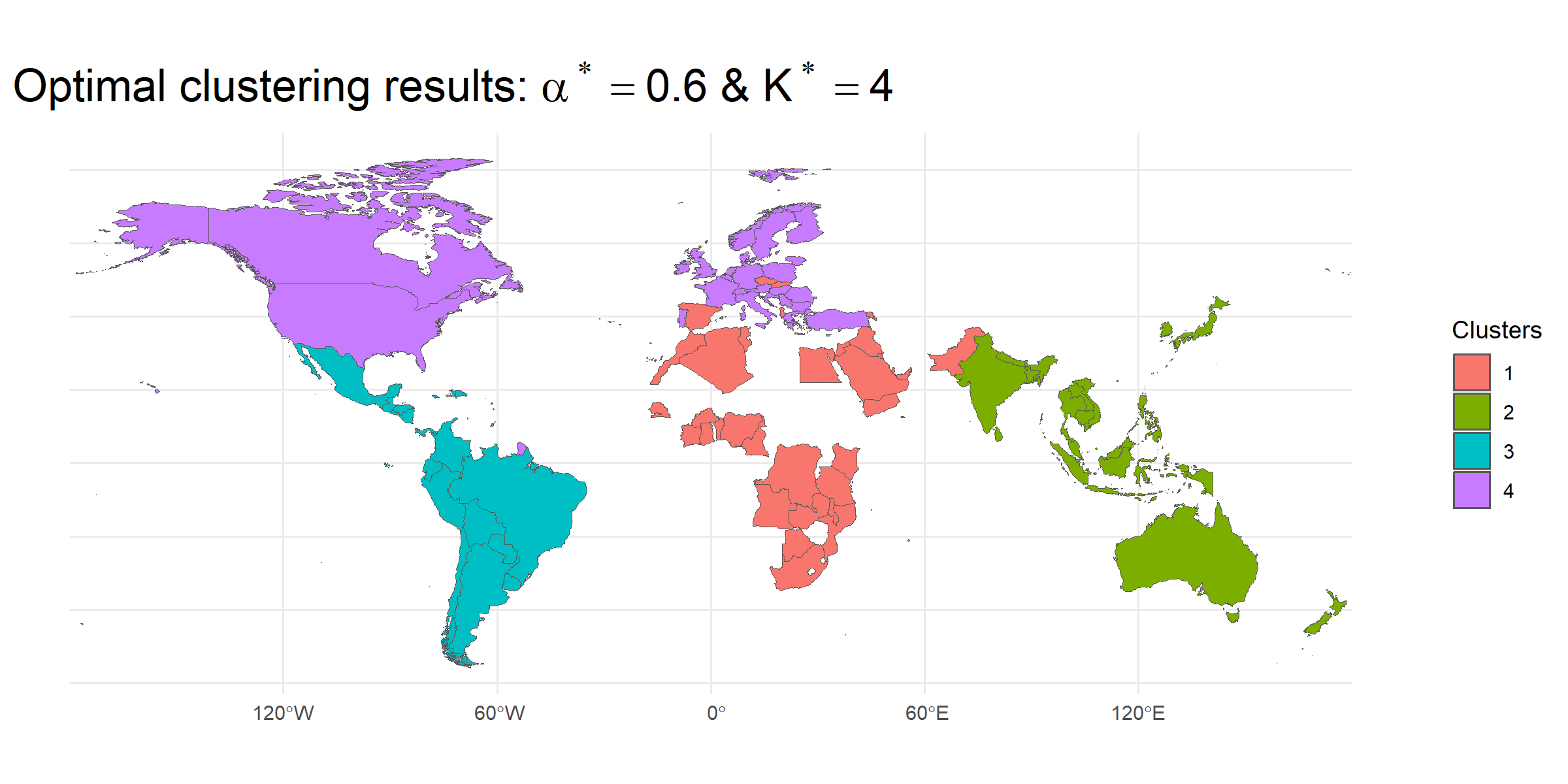}
	\caption{Map of clustering partitions obtained by setting $\alpha = 0.60$ and $K = 4$ and using as data the share of people in the 2022 survey with medium-low and low climate change awareness.}
	\label{fig:map_k4_a0.6}
\end{figure}

\begin{figure}
	\centering
	\includegraphics[width=1\columnwidth]{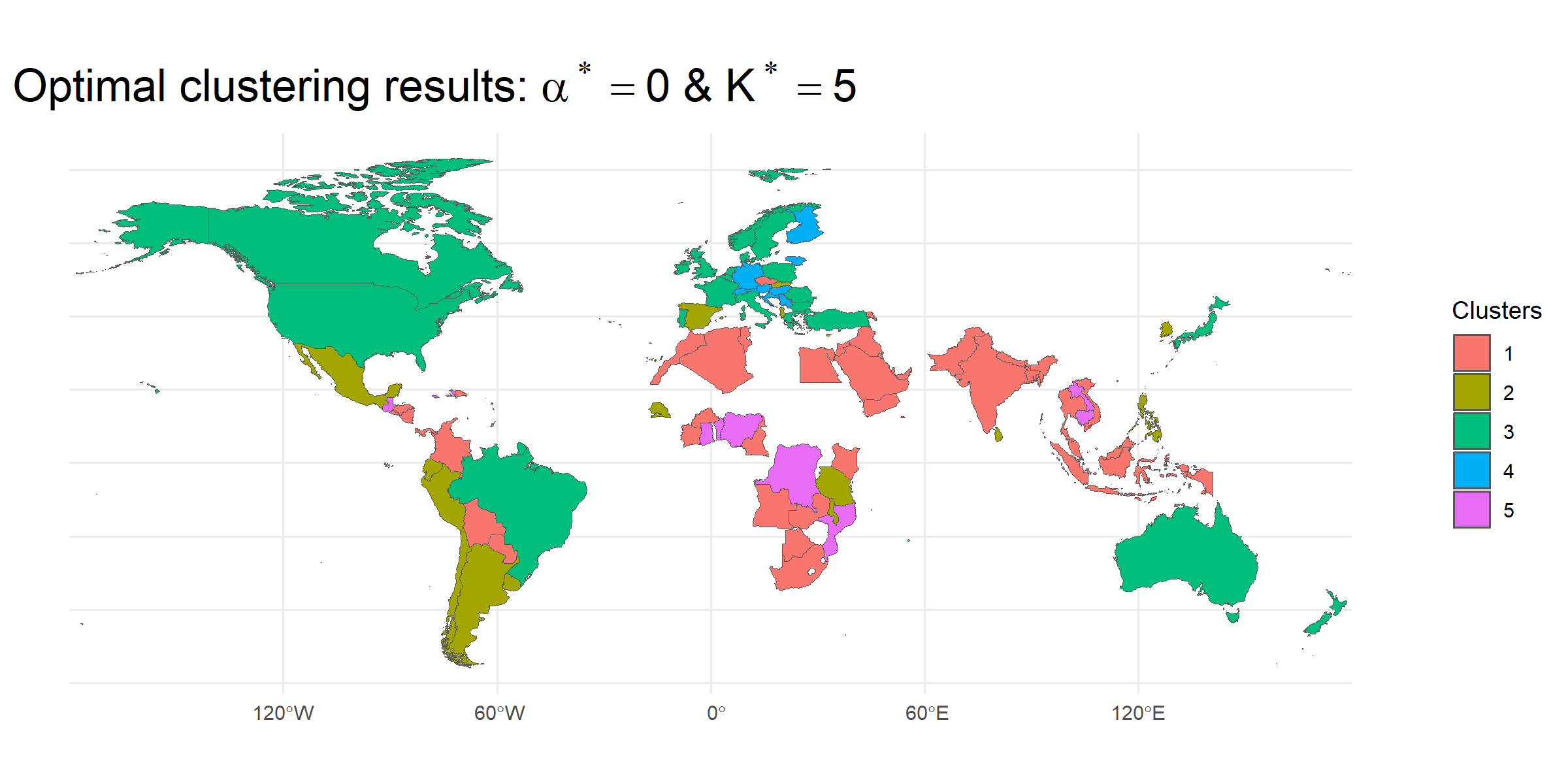}
	\caption{Map of clustering partitions obtained by setting $\alpha = 0$ and $K = 5$ and using as data the share of people in the 2022 survey with medium-low and low climate change awareness.}
	\label{fig:map_k5_a0}
\end{figure}

\begin{figure}
	\centering
	\includegraphics[width=1\columnwidth]{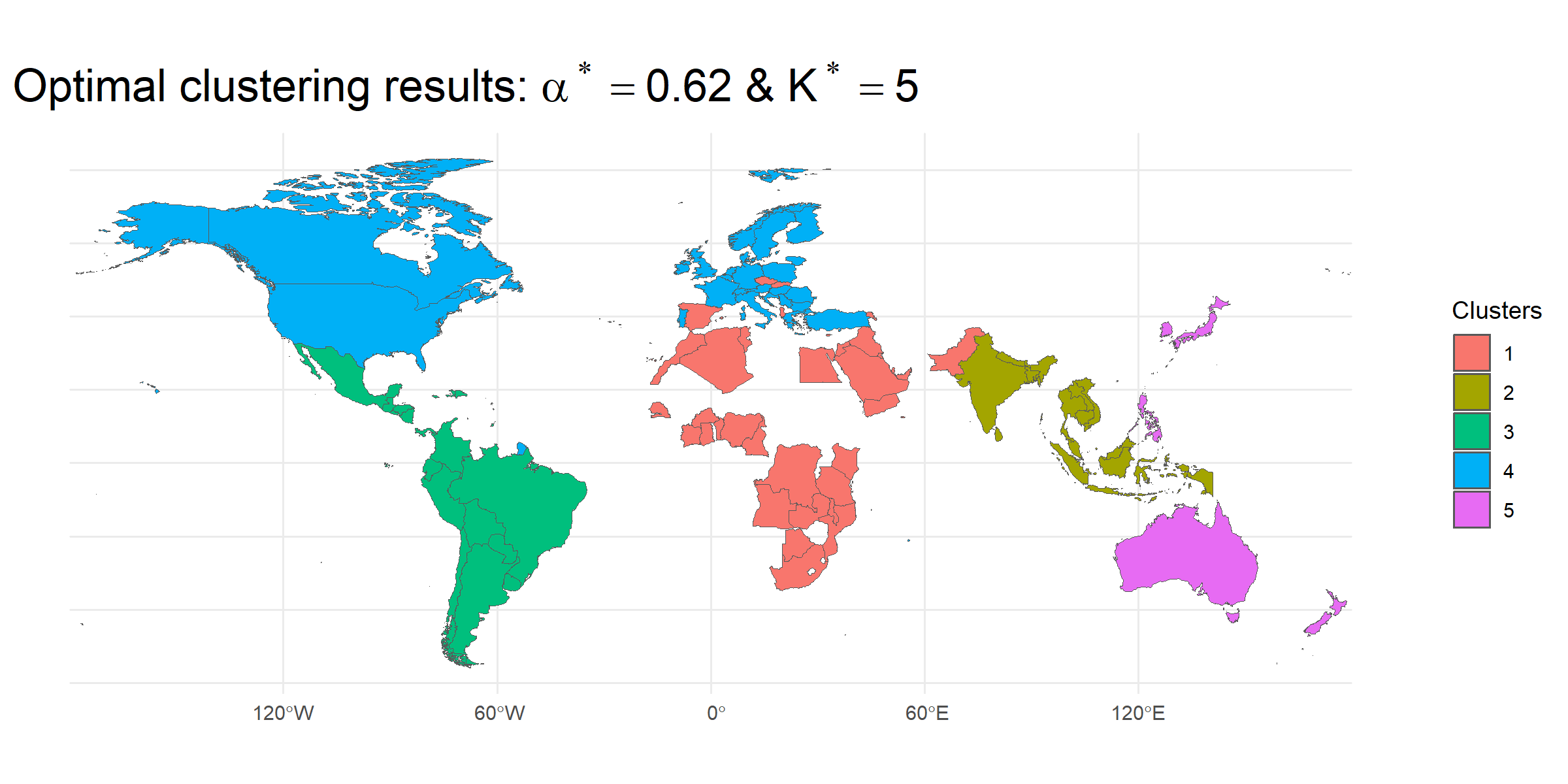}
	\caption{Map of clustering partitions obtained by setting $\alpha = 0.62$ and $K = 5$ and using as data the share of people in the 2022 survey with medium-low and low climate change awareness.}
	\label{fig:map_k5_a0.62}
\end{figure}

Figures \ref{fig:map_k4_a0} through \ref{fig:map_k5_a0.62} show two sets of maps produced for $K = 4$ and $K=5$ with $\alpha$ = 0 or $\alpha_{K,max}^* = 0.60$ (for $K=4$) or $\alpha_{K,max}^* = 0.62$ (for $K=5$). In this manner, a clustering with the features and the geographical components combined may be compared to a clustering without the geographic component restrictions. Figure \ref{fig:map_k4_a0.6} shows that the optimized $\alpha$ value produces a clustering that heavily considers the geographical component, but with some interesting exceptions such as the Czech Republic or Spain which, instead of being in the European cluster, are placed in the African countries cluster. This observation is consistent with what was observed in the comparison made in Figure \ref{fig:comparison21_22} which showed that Spain in 2022 transitioned to a cluster with lower awareness. In addition, Pakistan, which geographically should be placed in the Asian-Australian cluster, is instead located in the African countries cluster, while French Guiana instead of being in the South America cluster is in the North America cluster. With the addition of the fifth cluster (Figures \ref{fig:map_k5_a0} and \ref{fig:map_k5_a0.62}) some of the countries included in the Asian-Australian cluster such as Australia, New Zealand, the Philippines and Japan have split off and moved to an independent cluster.



\begin{figure}
	\centering
	\includegraphics[width=1\columnwidth]{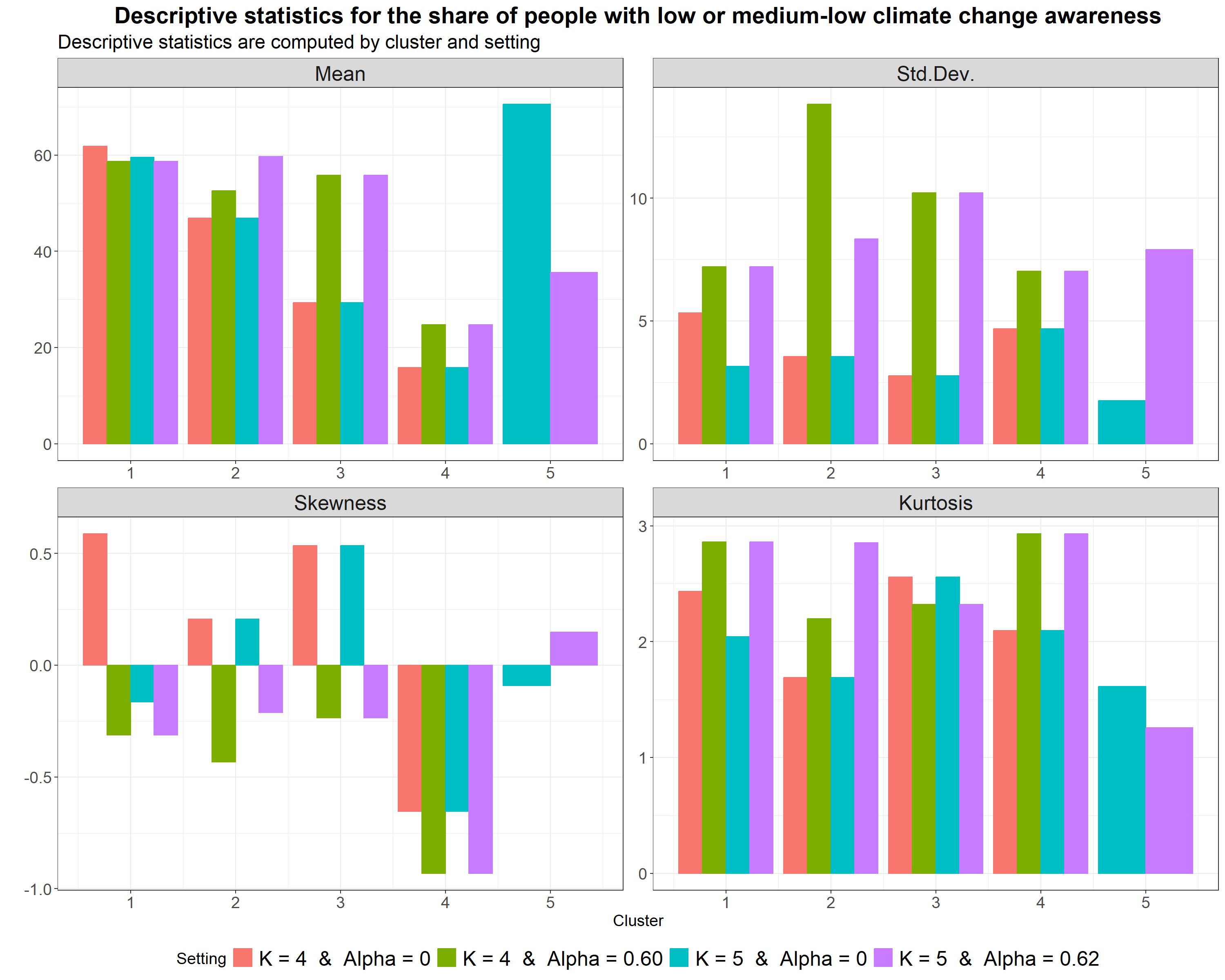}
	\caption{Descriptive statistics on the share of interviewed people declaring a low or medium-low climate change awareness. Statistics are computed by cluster and setting (i.e., combination of $\alpha$ and $K$). Colors allow to compare the statistics across clusters for different settings, while side-by-side bars allow the comparison of the statistics across the settings for a specific cluster.}
	\label{fig:descr_statistics}
\end{figure}

Figure \ref{fig:descr_statistics} shows the main descriptive statistics for each cluster across the considered settings. For instance, the first setting with $K = 4$ and $\alpha$ = 0 is represented by the red bars that show a decreasing average value of low awareness computed by taking into consideration the countries in each of the four clusters. In general, the four clusterings agree in joining European and North American countries and those with advanced Capitalist economies, as well as in joining African and Middle Eastern countries. Without particular surprise, these two blocs are well evidenced by the descriptive statistics, which show that awareness is significantly higher (on average) and more compact (reduced variability) in the Western block than in the other groups.

\subsection{Robustness analysis: hierarchical spatial clustering using climate change awareness, climate-related and socio-economic information}
\begin{figure}
	\centering
	\includegraphics[width=1\columnwidth]{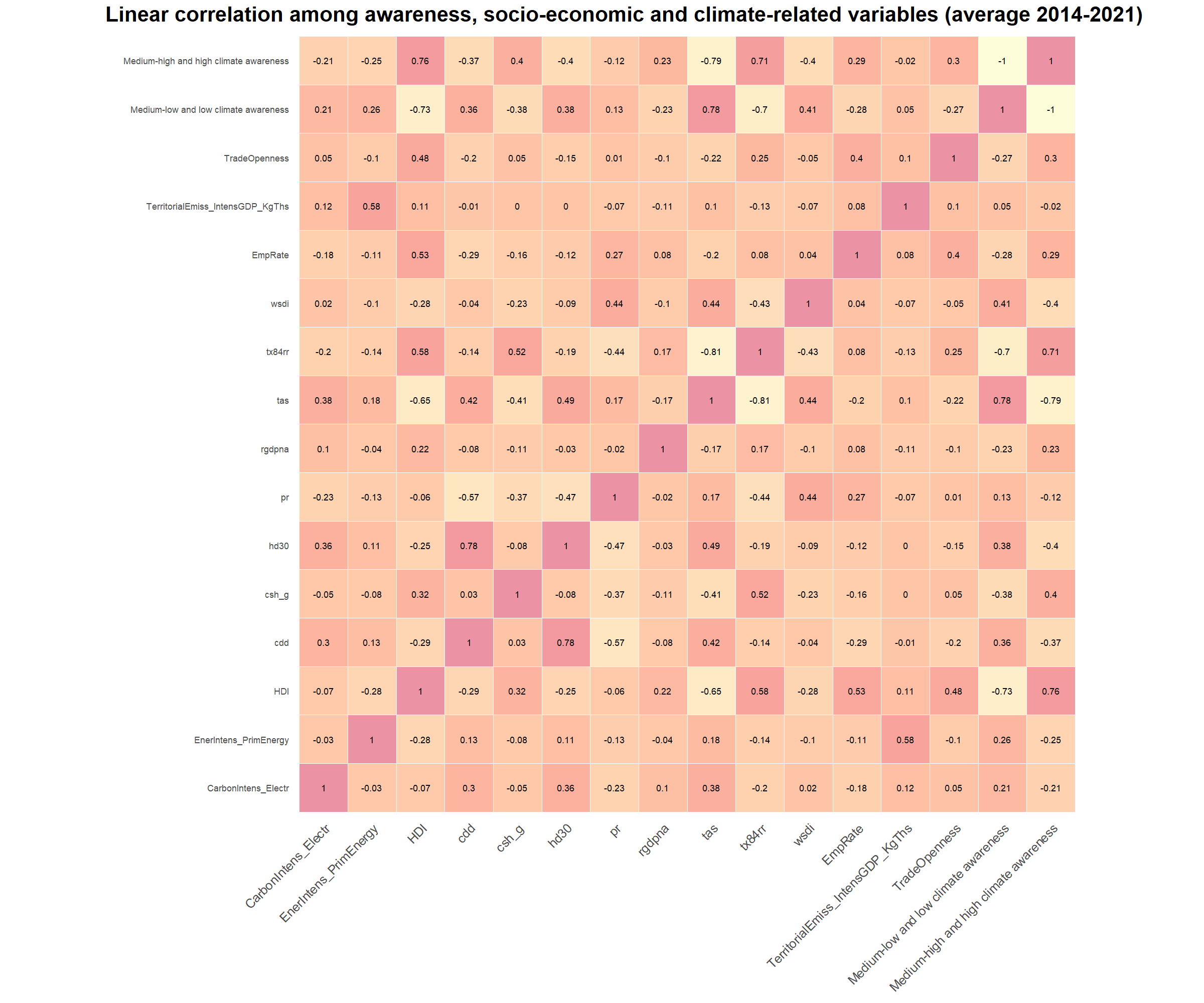}
	\caption{Heatmap of pairwise Pearson's linear correlation index for the full set of climate awareness, climate-related and socio-economic features.}
	\label{fig:heatmap}
\end{figure}

In the main analysis we built the feature dissimilarity matrix $D_0$ considering only the information provided the climate change awareness survey. In particular, we used the national share of interviewed people declaring a medium-low or low knowledge about climate change. To expand the analysis and use other variables that help us to better characterize the countries taken into consideration, we have collected the variables listed in Table \ref{tab:variables}. These variables are also shown in Figure \ref{fig:heatmap} where we computed the Pearson's linear correlation index. The medium-high and high awareness is positively correlated with a group of variables related to education (human capital index and human development index, years of schooling), the availability of basic services (e.g., electricity, sanitation, cooking technologies), and excess mortality risk (tx84rr). The medium-high and high awareness is also negatively correlated mainly only with the mean surface air temperature, while all the other negative correlations are relatively weak.

To strengthen the main findings we performed several robustness check in which we modified the set of features used to generate the clustering while letting unchanged the spatial dissimilarity matrix $D_1$. We considered four alternative scenarios:
\begin{enumerate}
    \item High and medium-high climate change awareness + geographical distance
    \item Low and medium-low climate change awareness + socio-economic features + geographical distance
    \item High and medium-high climate change awareness + climate-related features + geographical distance
    \item High and medium-high climate change awareness + socio-economic features + climate-related features + geographical distance
\end{enumerate}

The robustness analysis aims to check the coherence the partitions generated by the awareness only case (main analysis) with the alternative scenarios taking as fixed the number of clusters $K$ and the mixing parameter $\alpha$. This means that for all the alternative clusterings we used the values of $K$ and $\alpha$ identified in the main analysis, that is, $K^*=\{4,5\}$ and $\alpha=\{0,0.60,0.62\}$, and that we compare the subsequent clusters generated under different sets of features. The combination of these values provides a set of 12 alternative partitions. Given the large amount of potential outputs, we report the full set of results in the Appendix attached to the manuscript and in which the readers will find the maps corresponding to each scenario and combination of hyper-parameters. Here in the main text, we only provide a comparison of the resulting groups based on the Adjusted Rand Index or ARI (e.g., see Section 7.2 of \cite{Gordon1999}). Among the other, the ARI is used in classification and clustering to check if there exists a close correspondence between two independently-derived partitions of the same set of objects; and in particular, when the ARI computed on a pair of clustering results approaches 1 it indicates an high agreement among the two partitions, while negative values or near 0 suggest that the partitions do not agree on pairing.

\begin{figure}
	\centering
	\includegraphics[width=1\columnwidth]{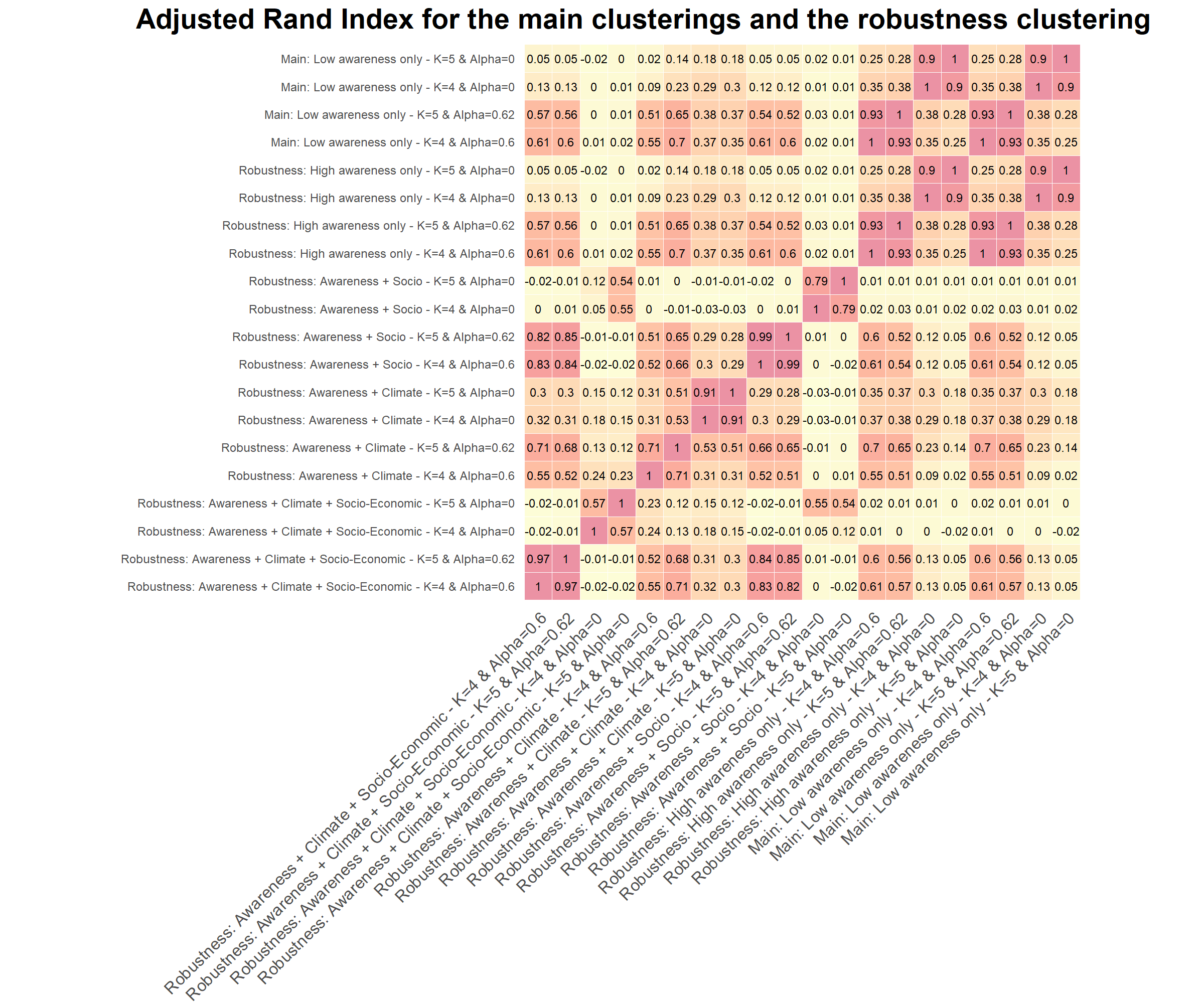}
	\caption{Heatmap of pairwise Adjusted Rand Index for the four clusterings from the main analysis (marked as \textit{main}) and the clusterings from the robustness checks. Values near to 1 indicate high similarity among the two partitions, while negative or null values (expected value is 0) is associated with partitions selected at random or completely different. Columns or rows marked as \textit{AwareClimateSocio} refer to the clustering jointly considering climate awareness, climate-related and socio-economic features; Columns or rows marked as \textit{AwareClimate} refer to the clustering jointly considering climate awareness and climate-related features; Columns or rows marked as \textit{AwareSocio} refer to the clustering jointly considering climate awareness and socio-economic features.}
	\label{fig:ari}
\end{figure}

The results from the robustness analysis can be summarized as follows:
\begin{itemize}
    \item As expected, using the share of respondents with high consciousness and that with low consciousness generates almost completely overlapping and consistent results (ARI above 0.90 for both cases of geographically-informed and purely-feature settings), showing the full interchangeability of the two quantities;
    \item The use of 4 or 5 groups generates highly coherent partitions, both in the case of geographically-informed and purely-feature clusterings: this is evident from the fact that the cells around the northeast-southwest diagonal have ARIs between 0.57 and 0.97;
    \item The results generated by the analyses in which geographic information is considered (rows 3, 4, 7 and 8 of the heatmap \ref{fig:ari}) are robust to the inclusion of socio-economic and climate change-related information, while they deviate strongly from the results obtained without spatial information (see, for example, intersections 3-3 and 4-4);
    \item Comparison of the maps in the Appendix shows that when geography is included, the clusters reflect (consistently) continental blocks (e.g., European countries are almost always clustered together, as is the Asian-Pacific block) or cultural-historical blocks (e.g., European countries, North America, and Australia). When geography is omitted some of the blocks remain (e.g., stability of European countries) while blocks such as South America and Africa are scattered and more heterogeneous. These considerations apply across the various scenarios considered;
\end{itemize}
In general, the above points suggest greater stability of partitions generated by including geographic information than the purely-feature clustering case. More specifically, the geographically-informed clustering leads to more geographically-compact aggregations (clusters reflect continental or cultural blocks); reduces the potential variations generated as the chosen number of clusters varies and thus to more stable and less volatile partitions; and generates consistent partitions even when information sets of different nature and varying numbers of features are considered.

\section{Conclusions}
\label{sec:conclusions}
In this paper we provide a hierarchical cluster analysis of the global climate change awareness at country level while accounting for spatial constraints. Besides the awareness, we collected a large data set that includes climate change awareness, climate-related and socio-economics variables. Our approach combines and extends the clustering methods proposed by \cite{chavent2018clustgeo} and by \cite{MorelliEtAl2024}, in which a mixing parameter $\alpha$ is used to linearly combine the dissimilarity matrix for the feature space (i.e., $D_0$) and the geographical dissimilarity (i.e., $D_1$). We propose a new algorithm to simultaneously optimize the values of $\alpha$ and the number of clusters $K$. Specifically, we propose to select the optimal combination of $\alpha^*_K$ and $K^*$ taking into account the within-clusters homogeneity, the between-clusters separation and a comparison between the geographically-informed and non-geographical clustering. We apply this method to the collected data set showing different scenarios in which countries are divided in clusters based on their level of climate change awareness subject to geographical constraints or on the extended data set of socio-economic and climate variables.

The results show that an optimization of $\alpha$ and $K$ using the proposed method allows to obtain a better results compared to a clustering where the geographical component is not considered (i.e., $\alpha$ = 0). In addition, robustness analyses conducted implementing different sets of socio-economic and climate-related information allowed to obtain results consistent with the main analysis. In particular, we found that geographically-informed clustering guarantees a greater stability of partitions with respect to the the purely-feature clustering case, as well as it allows for a greater geographical compactness of groups, which often reflect continental or cultural blocks of countries in the World. Indeed, the identified clusters confirm the presence of a clear dualism in terms of awareness about climate change in which Western countries (Europe, America, and Australian countries) are contrasted with Asian, African, and Middle Eastern countries. While awareness seems high and compact in the former, there is greater variability but still lower awareness in the latter.

Future developments of this work will include the development of a method to predict the level of climate change awareness of countries currently not included in the survey, for which awareness might be approximated based on the cluster assigned according to the non-awareness (socio-economic, climate, and geographical) variables.



\section*{Declarations}
\subsection*{Conflict of interest}
Conflict of interest/Competing interests: The authors have no competing interests to declare that are relevant to the content of this article. Also, the authors did not receive support from any organization for the submitted work;
\subsection*{Data availability}
Data availability and codes: All results presented in this paper can be reproduced using the R software. The codes were developed entirely by the authors. Attached to the submission files, we attach the complete set of data set and scripts. For the reproducibility purposes, all scripts and the data are made available for the public on the following GitHub folder: \href{https://github.com/PaoloMaranzano/GPZ\_PM\_HierSpatClusterClimateAwareness.git}{https://github.com/PaoloMaranzano/GPZ\_PM\_HierSpatClusterClimateAwareness.git}


\bibliographystyle{unsrt}  
\bibliography{cas-refs.bib}

\begin{thebibliography}{10}

\bibitem{loarie2009velocity}
Scott~R Loarie, Philip~B Duffy, Healy Hamilton, Gregory~P Asner, Christopher~B
  Field, and David~D Ackerly.
\newblock The velocity of climate change.
\newblock {\em Nature}, 462(7276):1052--1055, 2009.

\bibitem{wang2023climate}
Fang Wang, Jean~Damascene Harindintwali, Ke~Wei, Yuli Shan, Zhifu Mi, Mark~John
  Costello, Sabine Grunwald, Zhaozhong Feng, Faming Wang, Yuming Guo, et~al.
\newblock Climate change: Strategies for mitigation and adaptation.
\newblock {\em The Innovation Geoscience}, 1(1):100015--61, 2023.

\bibitem{canadell2021multi}
Josep~G Canadell, CP~Meyer, Garry~D Cook, Andrew Dowdy, Peter~R Briggs,
  J{\"u}rgen Knauer, Acacia Pepler, and Vanessa Haverd.
\newblock Multi-decadal increase of forest burned area in australia is linked
  to climate change.
\newblock {\em Nature communications}, 12(1):6921, 2021.

\bibitem{goss2020climate}
Michael Goss, Daniel~L Swain, John~T Abatzoglou, Ali Sarhadi, Crystal~A Kolden,
  A~Park Williams, and Noah~S Diffenbaugh.
\newblock Climate change is increasing the likelihood of extreme autumn
  wildfire conditions across california.
\newblock {\em Environmental Research Letters}, 15(9):094016, 2020.

\bibitem{sun2022understanding}
Ying Sun, Xuebin Zhang, Yihui Ding, Deliang Chen, Dahe Qin, and Panmao Zhai.
\newblock Understanding human influence on climate change in china.
\newblock {\em National science review}, 9(3):nwab113, 2022.

\bibitem{meza2021drought}
Isabel Meza, Ehsan~Eyshi Rezaei, Stefan Siebert, Gohar Ghazaryan, Hamideh
  Nouri, Olena Dubovyk, Helena Gerdener, Claudia Herbert, J{\"u}rgen Kusche,
  Eklavyya Popat, et~al.
\newblock Drought risk for agricultural systems in south africa: Drivers,
  spatial patterns, and implications for drought risk management.
\newblock {\em Science of the Total Environment}, 799:149505, 2021.

\bibitem{roman2020recent}
Cristian Rom{\'a}n-Palacios and John~J Wiens.
\newblock Recent responses to climate change reveal the drivers of species
  extinction and survival.
\newblock {\em Proceedings of the National Academy of Sciences},
  117(8):4211--4217, 2020.

\bibitem{IPBES2019}
Purvis Andy.
\newblock A million threatened species? thirteen questions and answers, 2019.
\newblock
  \url{https://www.ipbes.net/news/million-threatened-species-thirteen-questions-answers},
  Last accessed on August 31 2024.

\bibitem{titus1986greenhouse}
James~G Titus.
\newblock Greenhouse effect, sea level rise, and coastal zone management.
\newblock {\em Coastal Management}, 14(3):147--171, 1986.

\bibitem{zahran2007ecological}
Sammy Zahran, Eunyi Kim, Xi~Chen, and Mark Lubell.
\newblock Ecological development and global climate change: A cross-national
  study of kyoto protocol ratification.
\newblock {\em Society and Natural Resources}, 20(1):37--55, 2007.

\bibitem{edmonds2003costs}
James~A Edmonds and Ronald~D Sands.
\newblock What are the costs of limiting co2 concentrations.
\newblock {\em Global climate change: The science, economics, and politics},
  pages 140--86, 2003.

\bibitem{GCRI2021}
David Eckstein, Vera Künzel, and Schäfer Laura.
\newblock Global climate risk index, 2021.
\newblock
  \url{https://reliefweb.int/report/world/global-climate-risk-index-2021}, Last
  accessed on August 31 2024.

\bibitem{gordon1996survey}
AD~Gordon.
\newblock A survey of constrained classification.
\newblock {\em Computational Statistics \& Data Analysis}, 21(1):17--29, 1996.

\bibitem{ambroise1997clustering}
Christophe Ambroise, Mo~Dang, and G{\'e}rard Govaert.
\newblock Clustering of spatial data by the em algorithm.
\newblock In {\em geoENV I—Geostatistics for Environmental Applications:
  Proceedings of the Geostatistics for Environmental Applications Workshop,
  Lisbon, Portugal, 18--19 November 1996}, pages 493--504. Springer, 1997.

\bibitem{liao2012clustering}
Zhung-Xun Liao and Wen-Chih Peng.
\newblock Clustering spatial data with a geographic constraint: exploring local
  search.
\newblock {\em Knowledge and information systems}, 31:153--170, 2012.

\bibitem{miele2014spatially}
Vincent Miele, Franck Picard, and St{\'e}phane Dray.
\newblock Spatially constrained clustering of ecological networks.
\newblock {\em Methods in Ecology and Evolution}, 5(8):771--779, 2014.

\bibitem{pawitan2003constrained}
Yudi Pawitan and Jian Huang.
\newblock Constrained clustering of irregularly sampled spatial data.
\newblock {\em Journal of Statistical Computation and Simulation},
  73(12):853--865, 2003.

\bibitem{oliver1989geostatistical}
MA~Oliver and R~Webster.
\newblock A geostatistical basis for spatial weighting in multivariate
  classification.
\newblock {\em Mathematical geology}, 21:15--35, 1989.

\bibitem{bourgault1992multivariate}
Gilles Bourgault, Denis Marcotte, and Pierre Legendre.
\newblock The multivariate (co) variogram as a spatial weighting function in
  classification methods.
\newblock {\em Mathematical Geology}, 24:463--478, 1992.

\bibitem{chavent2018clustgeo}
Marie Chavent, Vanessa Kuentz-Simonet, Amaury Labenne, and J{\'e}r{\^o}me
  Saracco.
\newblock Clustgeo: an r package for hierarchical clustering with spatial
  constraints.
\newblock {\em Computational Statistics}, 33(4):1799--1822, 2018.

\bibitem{leiserowitz2021international}
Anthony Leiserowitz, Jennifer Carman, Nicole Buttermore, Xinran Wang, Seth
  Rosenthal, Jennifer Marlon, and Kelsey Mulcahy.
\newblock International public opinion on climate change.
\newblock {\em New Haven, CT: Yale Program on Climate Change Communication and
  Facebook Data for Good}, 2021.

\bibitem{leiserowitz2022international}
Anthony Leiserowitz, Jennifer Carman, Nicole Buttermore, Liz Neyens, Seth
  Rosenthal, Jennifer Marlon, JW~Schneider, and Kelsey Mulcahy.
\newblock International public opinion on climate change 2022.
\newblock {\em New Haven, CT: Yale Program on Climate Change Communication and
  Facebook Data for Good}, 2022.

\bibitem{CRAN}
{R Core Team}.
\newblock {\em R: A Language and Environment for Statistical Computing}.
\newblock R Foundation for Statistical Computing, Vienna, Austria, 2023.

\bibitem{dendextend}
Tal Galili.
\newblock dendextend: an r package for visualizing, adjusting, and comparing
  trees of hierarchical clustering.
\newblock {\em Bioinformatics}, 2015.

\bibitem{AkhanliHennig2020}
Serhat~Emre Akhanli and Christian Hennig.
\newblock Comparing clusterings and numbers of clusters by aggregation of
  calibrated clustering validity indexes.
\newblock {\em Statistics and Computing}, 30(5):1523--1544, 2020.

\bibitem{RossbroichEtAl2022}
Julian Rossbroich, Jeffrey Durieux, and Tom~F. Wilderjans.
\newblock Model selection strategies for determining the optimal number of
  overlapping clusters in additive overlapping partitional clustering.
\newblock {\em Journal of Classification}, 39(2):264--301, 2022.

\bibitem{MorelliEtAl2024}
Caterina Morelli, Simone Boccaletti, Paolo Maranzano, and Philipp Otto.
\newblock Multidimensional spatiotemporal clustering--an application to
  environmental sustainability scores in europe.
\newblock {\em arXiv preprint arXiv:2405.20191}, 2024.

\bibitem{NbClust2014}
Malika Charrad, Nadia Ghazzali, Véronique Boiteau, and Azam Niknafs.
\newblock Nbclust: An r package for determining the relevant number of clusters
  in a data set.
\newblock {\em Journal of Statistical Software}, 61(6):1 -- 36, 2014.

\bibitem{Rousseeuw1987}
Peter~J Rousseeuw.
\newblock Silhouettes: a graphical aid to the interpretation and validation of
  cluster analysis.
\newblock {\em Journal of computational and applied mathematics}, 20:53--65,
  1987.

\bibitem{Dunn1974}
Joseph~C Dunn.
\newblock Well-separated clusters and optimal fuzzy partitions.
\newblock {\em Journal of cybernetics}, 4(1):95--104, 1974.

\bibitem{HubertLevin1976}
Lawrence~J Hubert and Joel~R Levin.
\newblock A general statistical framework for assessing categorical clustering
  in free recall.
\newblock {\em Psychological bulletin}, 83(6):1072, 1976.

\bibitem{CH1974}
Tadeusz Caliński and Jerzy Harabasz.
\newblock A dendrite method for cluster analysis.
\newblock {\em Communications in Statistics-theory and Methods}, 3(1):1--27,
  1974.

\bibitem{McClainRao1975}
John~O McClain and Vithala~R Rao.
\newblock Clustisz: A program to test for the quality of clustering of a set of
  objects.
\newblock {\em Journal of Marketing Research}, pages 456--460, 1975.

\bibitem{Gordon1999}
Allan~David Gordon.
\newblock {\em Classification}.
\newblock CRC Press, 1999.

\end{thebibliography}

\newpage

\appendix

\clearpage
\section{Main analysis results: spatial hierarchical clustering using low and medium-low climate change awareness information}
\begin{figure}[!htbp]
	\centering
	\includegraphics[width=1\columnwidth]{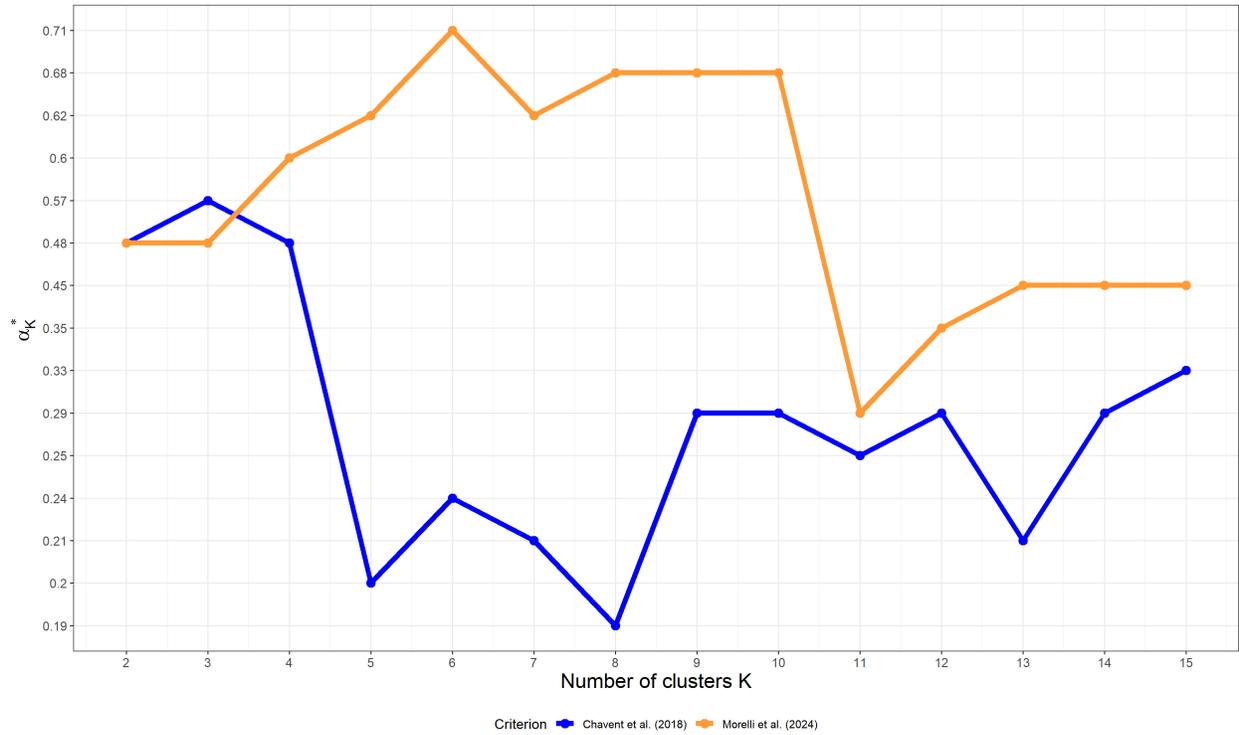}
	\caption{Optimal values of the mixing parameter $\alpha$ conditioning on $K$ clusters (that is, $\alpha_K^*$) obtained using the criterion by Morelli et al. (2024) ($\alpha_{K,max}^*$ in orange) and by Chavent et al. (2018) ($\alpha_{K,min}^*$ in blue).}
	\label{figA:optim_alpha}
\end{figure}

\begin{figure}[!htbp]
	\centering
	\includegraphics[width=1\columnwidth]{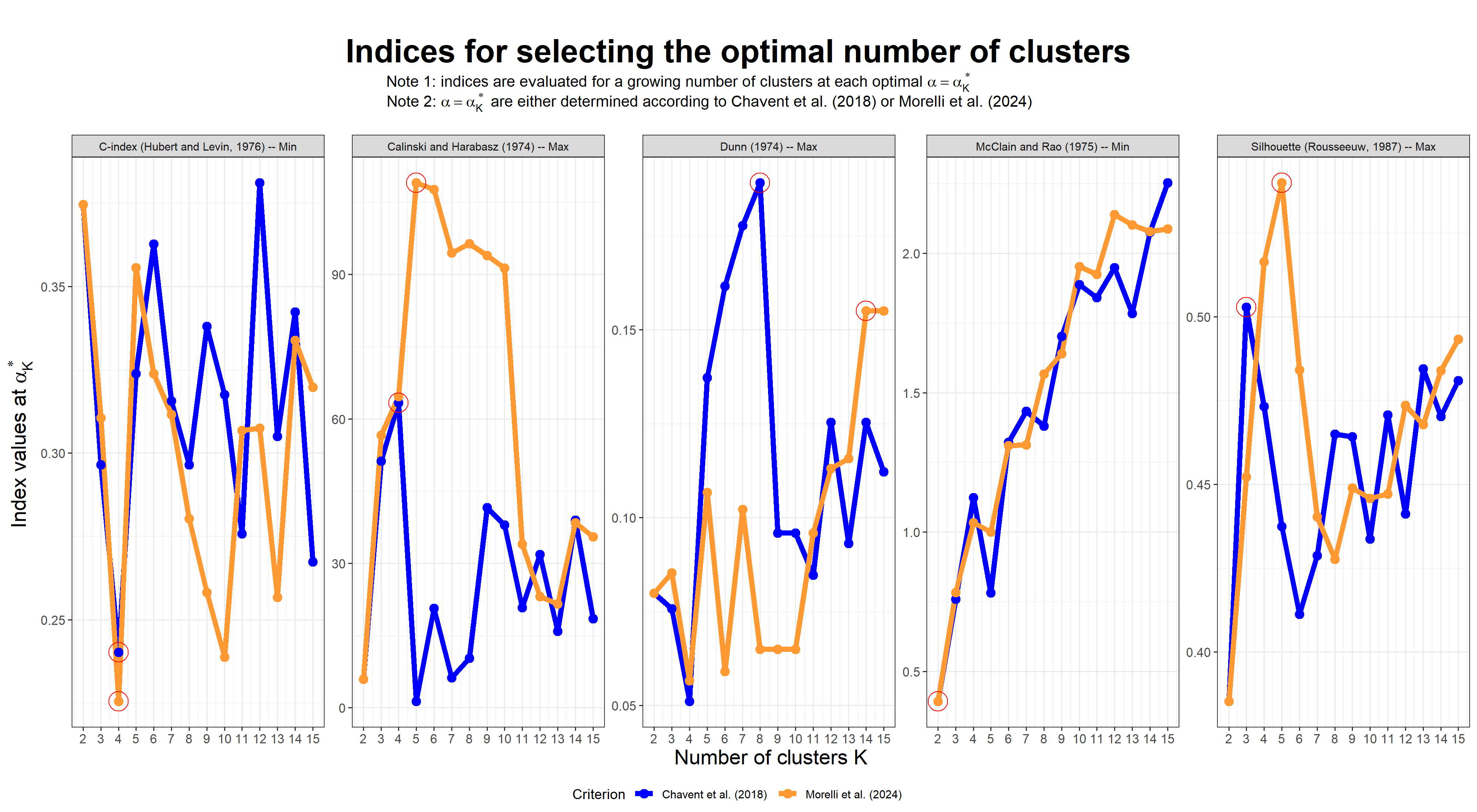}
	\caption{Estimates of the Silhouette index, the Dunn's index, the C-index, the Calinski-Harabasz's index, and the McClain-Rao's index for $K$ clusters evaluated at the corresponding $\alpha_{K,max}^*$ (orange) and $\alpha_{K,min}^*$ (blue). Red circles identifies the optimal values, that is, the minimizer for C-index and McClain-Rao's indices and the maximizer for Silhouette, Dunn's and Calinski-Harabasz's indices.}
	\label{figA:indices_optimK}
\end{figure}

\begin{figure}[!htbp]
	\centering
	\includegraphics[width=1\columnwidth]{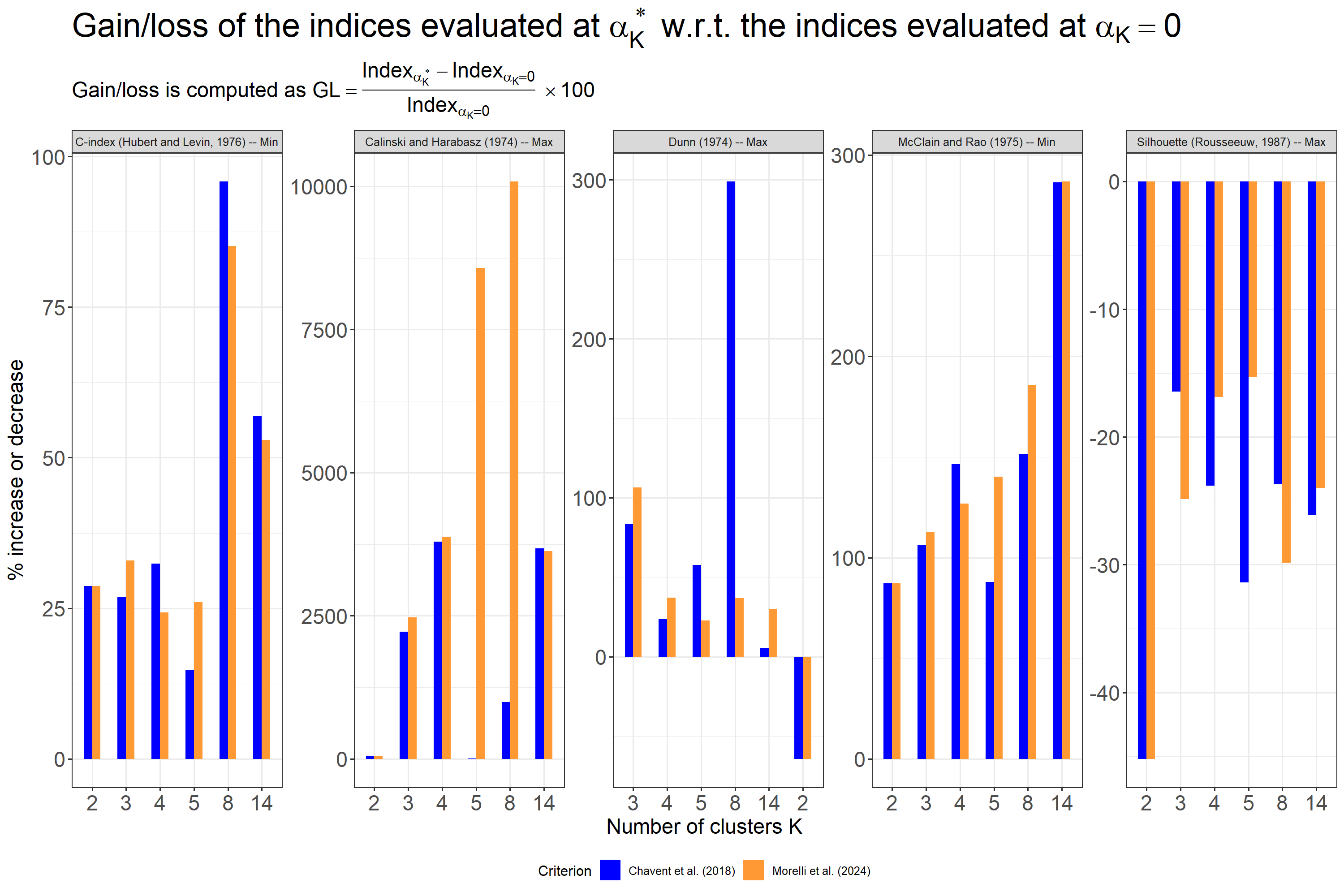}
	\caption{Estimated percentage gain (positive values) or loss (negative values) of the Silhouette, Dunn's, C-index, Calinski-Harabasz's index, and McClain-Rao's indices evaluated at the corresponding $\alpha_{K,max}^*$ (orange) and $\alpha_{K,min}^*$ (blue) with respect to the indices evaluated at $\alpha = 0$. Percentage gain/loss is computed as $GL = \frac{Index_{\alpha^*_K} - Index_{\alpha_K = 0}}{Index_{\alpha_K = 0}} \times 100$; thus, positive values have to be read as the percentage increase of the indices w.r.t. the baseline case of index at $\alpha = 0$, while negative values have to be read as the percentage decrease of the indices w.r.t. the baseline case of index at $\alpha = 0$.}
	\label{figA:gain_loss_alpha0}
\end{figure}

\begin{figure}[!htbp]
	\centering
	\includegraphics[width=1\columnwidth]{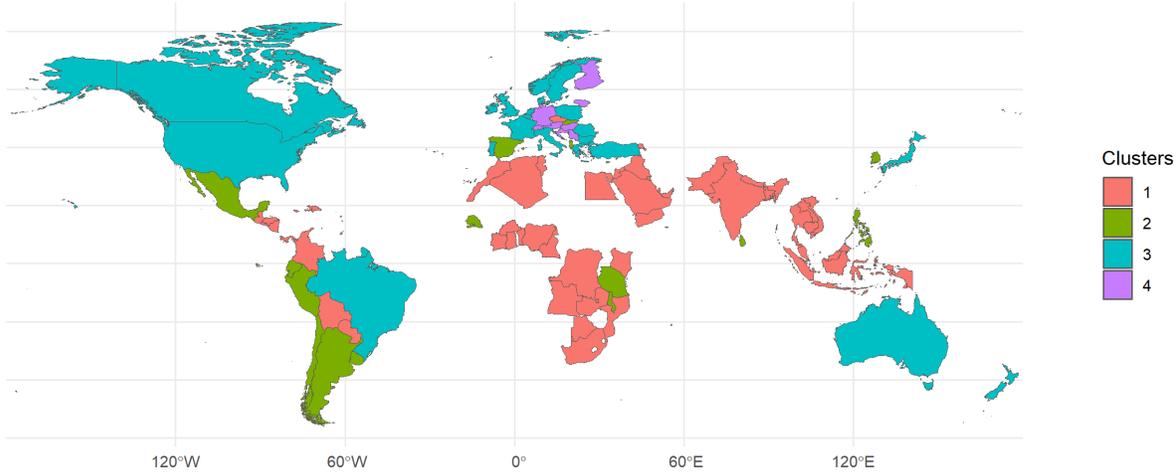}
	\caption{Map of clustering partitions obtained by setting $\alpha = 0$ and $K = 4$ and using as data the share of people in the 2022 survey with medium-low and low climate change awareness.}
	\label{figA:map_k4_a0}
\end{figure}

\begin{figure}[!htbp]
	\centering
	\includegraphics[width=1\columnwidth]{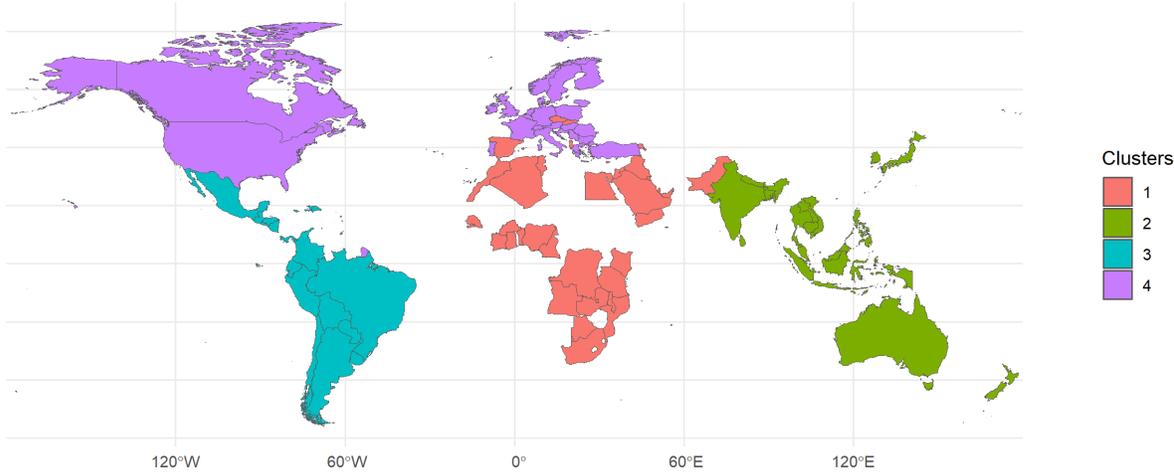}
	\caption{Map of clustering partitions obtained by setting $\alpha = 0.60$ and $K = 4$ and using as data the share of people in the 2022 survey with medium-low and low climate change awareness.}
	\label{figA:map_k4_a0.6}
\end{figure}

\begin{figure}[!htbp]
	\centering
	\includegraphics[width=1\columnwidth]{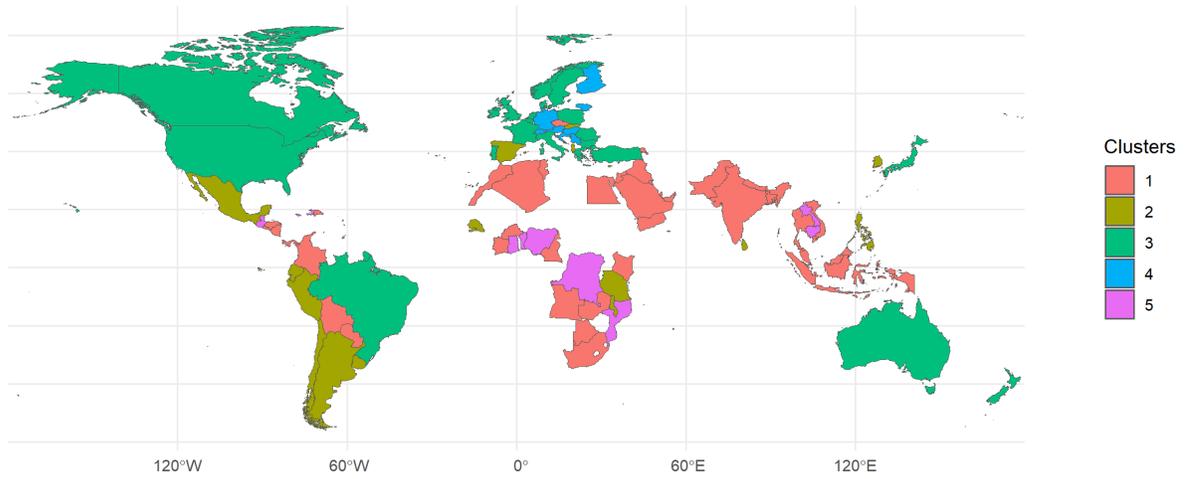}
	\caption{Map of clustering partitions obtained by setting $\alpha = 0$ and $K = 5$ and using as data the share of people in the 2022 survey with medium-low and low climate change awareness.}
	\label{figA:map_k5_a0}
\end{figure}

\begin{figure}[!htbp]
	\centering
	\includegraphics[width=1\columnwidth]{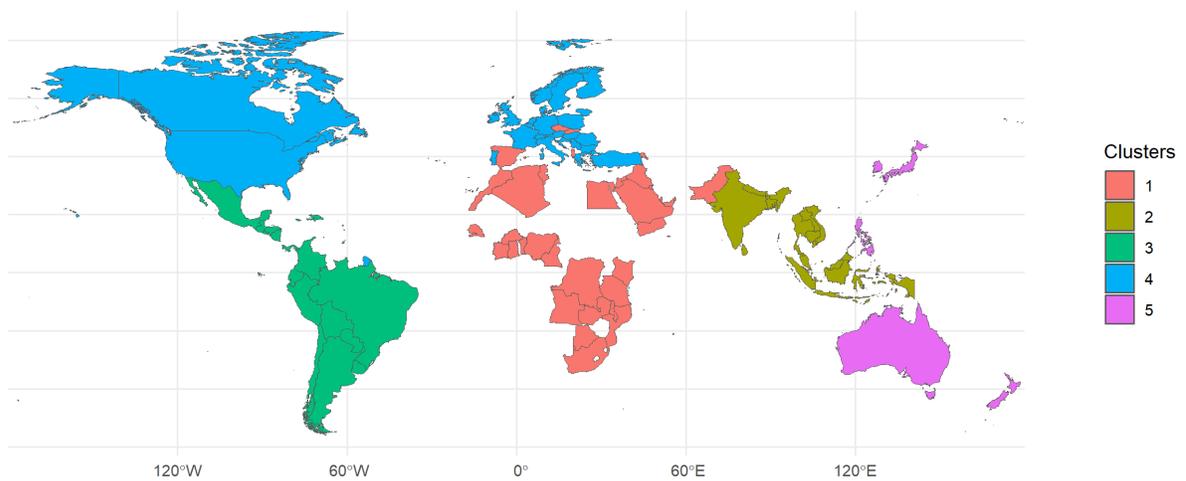}
	\caption{Map of clustering partitions obtained by setting $\alpha = 0.62$ and $K = 5$ and using as data the share of people in the 2022 survey with medium-low and low climate change awareness.}
	\label{figA:map_k5_a0.62}
\end{figure}

\begin{figure}[!htbp]
	\centering
	\includegraphics[width=1\columnwidth]{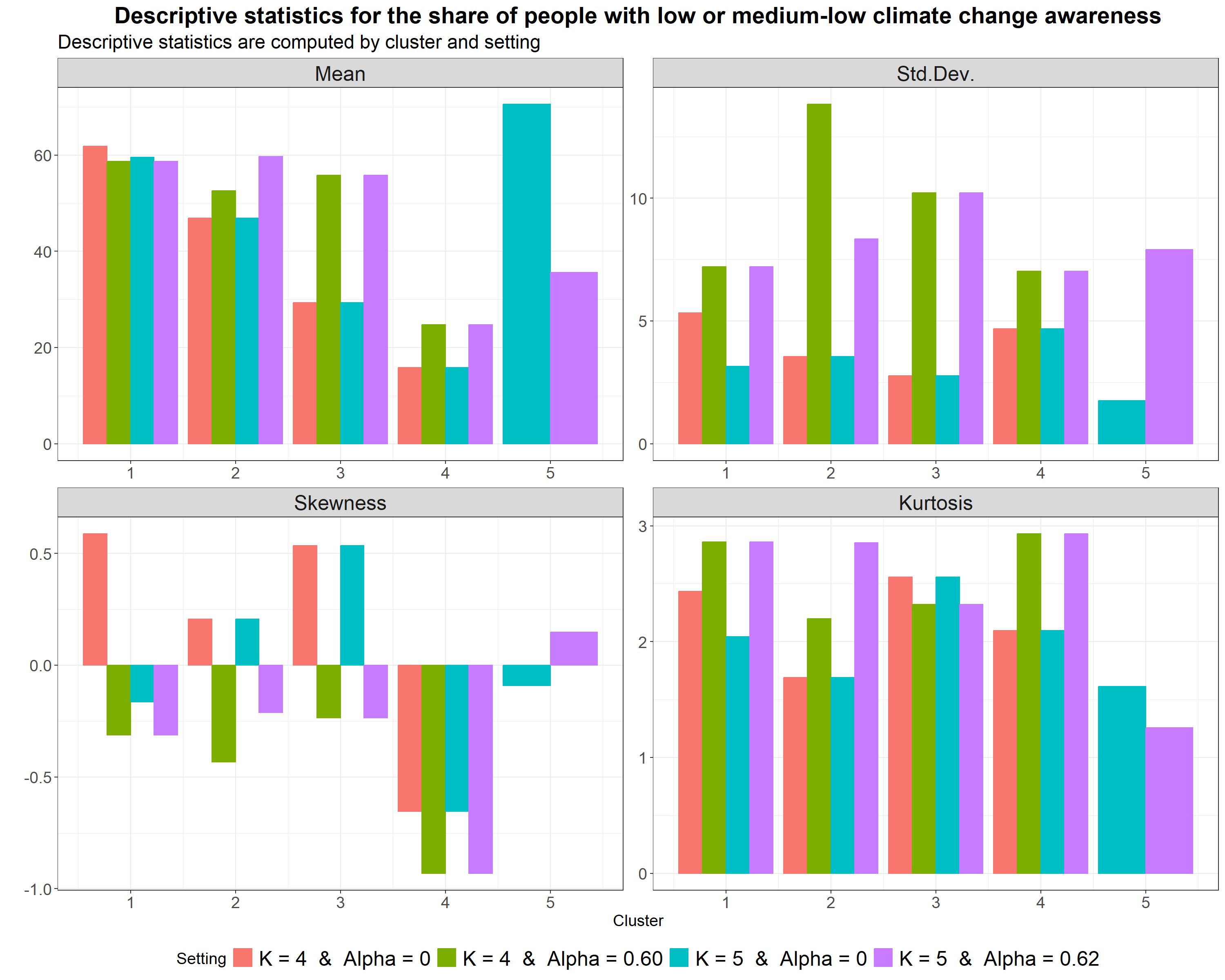}
	\caption{Descriptive statistics on the share of interviewed people declaring a low or medium-low climate change awareness. Statistics are computed by cluster and setting (i.e., combination of $\alpha$ and $K$). Colors allow to compare the statistics across clusters for different settings, while side-by-side bars allow the comparison of the statistics across the settings for a specific cluster.}
	\label{figA:descr_statistics}
\end{figure}

\clearpage
\section{Robustness analysis 1: spatial hierarchical clustering using high and medium-high climate change awareness information}
\begin{figure}[!htbp]
	\centering
	\includegraphics[width=1\columnwidth]{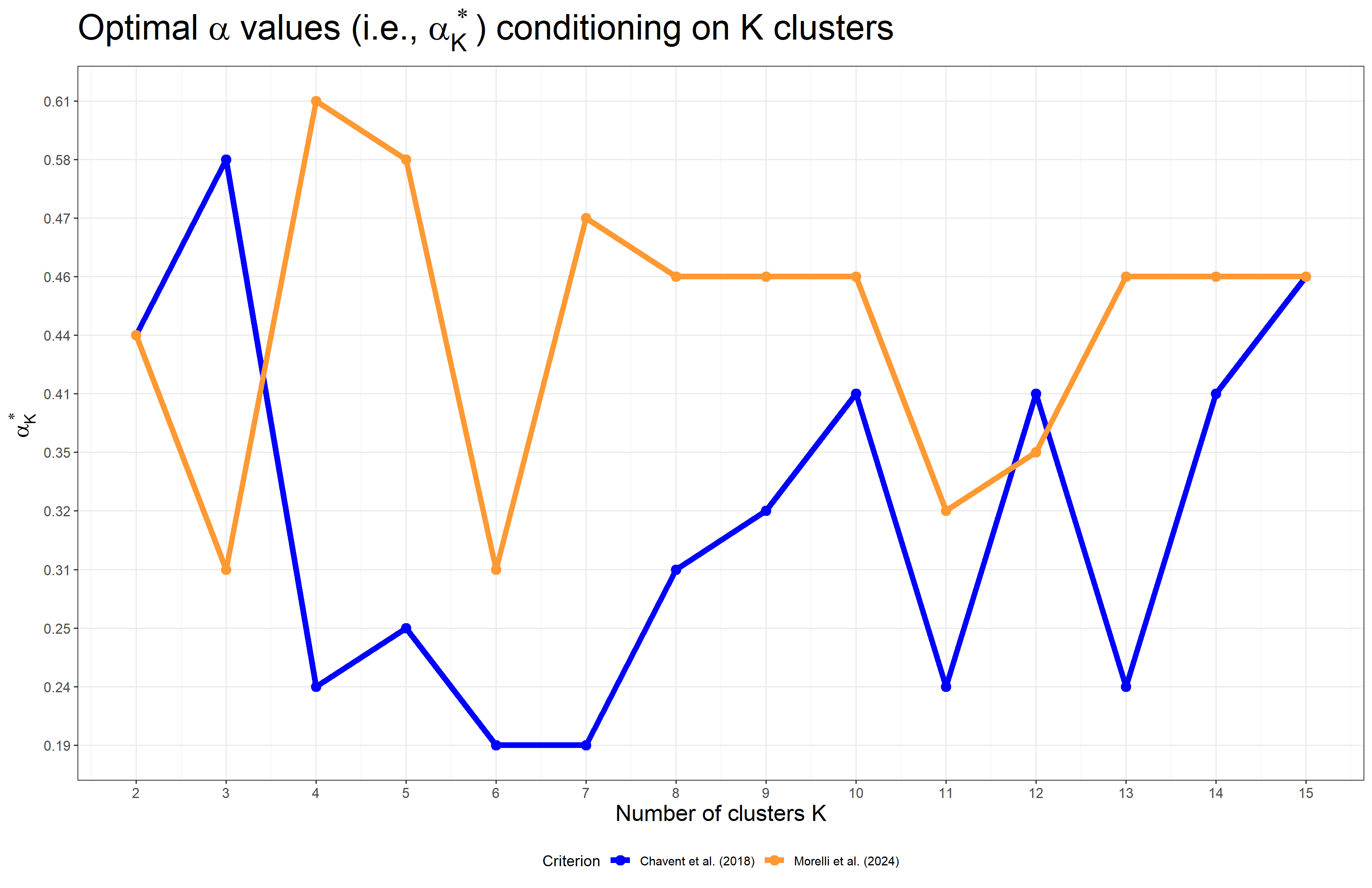}
	\caption{Optimal values of the mixing parameter $\alpha$ conditioning on $K$ clusters (that is, $\alpha_K^*$) obtained using the criterion by Morelli et al. (2024) ($\alpha_{K,max}^*$ in orange) and by Chavent et al. (2018) ($\alpha_{K,min}^*$ in blue).}
	\label{figB:optim_alpha}
\end{figure}

\begin{figure}[!htbp]
	\centering
	\includegraphics[width=1\columnwidth]{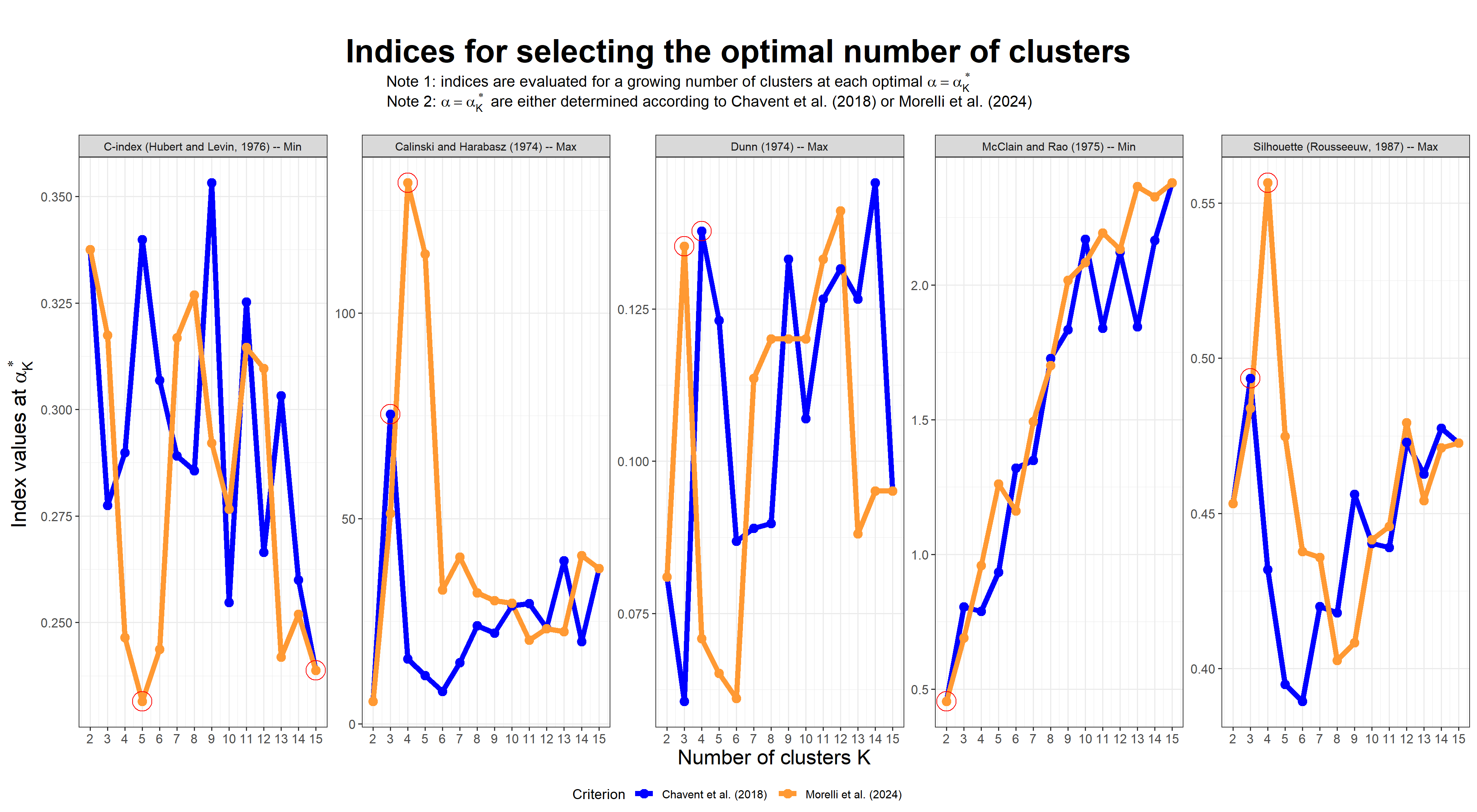}
	\caption{Estimates of the Silhouette index, the Dunn's index, the C-index, the Calinski-Harabasz's index, and the McClain-Rao's index for $K$ clusters evaluated at the corresponding $\alpha_{K,max}^*$ (orange) and $\alpha_{K,min}^*$ (blue). Red circles identifies the optimal values, that is, the minimizer for C-index and McClain-Rao's indices and the maximizer for Silhouette, Dunn's and Calinski-Harabasz's indices.}
	\label{figB:indices_optimK}
\end{figure}

\begin{figure}[!htbp]
	\centering
	\includegraphics[width=1\columnwidth]{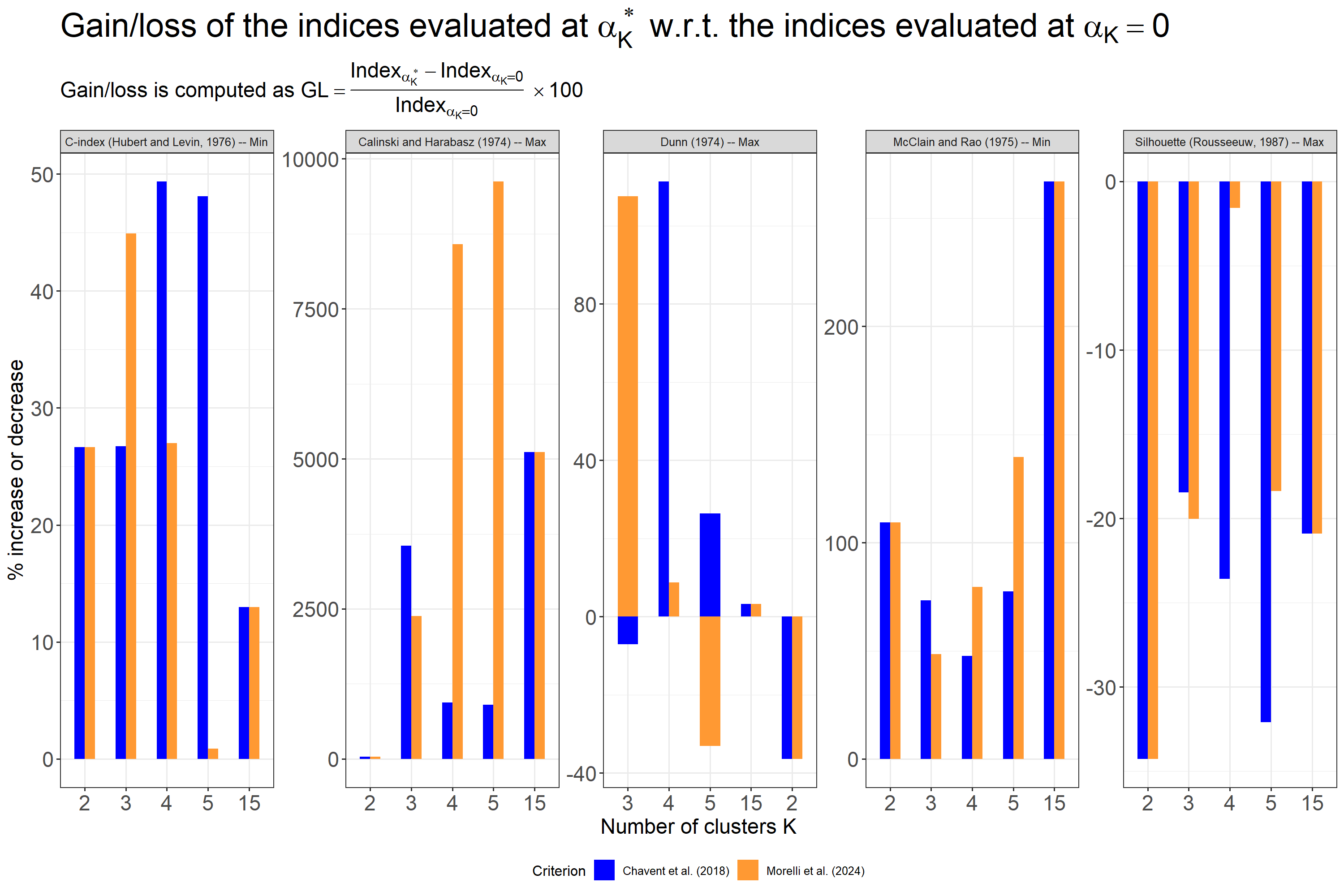}
	\caption{Estimated percentage gain (positive values) or loss (negative values) of the Silhouette, Dunn's, C-index, Calinski-Harabasz's index, and McClain-Rao's indices evaluated at the corresponding $\alpha_{K,max}^*$ (orange) and $\alpha_{K,min}^*$ (blue) with respect to the indices evaluated at $\alpha = 0$. Percentage gain/loss is computed as $GL = \frac{Index_{\alpha^*_K} - Index_{\alpha_K = 0}}{Index_{\alpha_K = 0}} \times 100$; thus, positive values have to be read as the percentage increase of the indices w.r.t. the baseline case of index at $\alpha = 0$, while negative values have to be read as the percentage decrease of the indices w.r.t. the baseline case of index at $\alpha = 0$.}
	\label{figB:gain_loss_alpha0}
\end{figure}

\begin{figure}[!htbp]
	\centering
	\includegraphics[width=1\columnwidth]{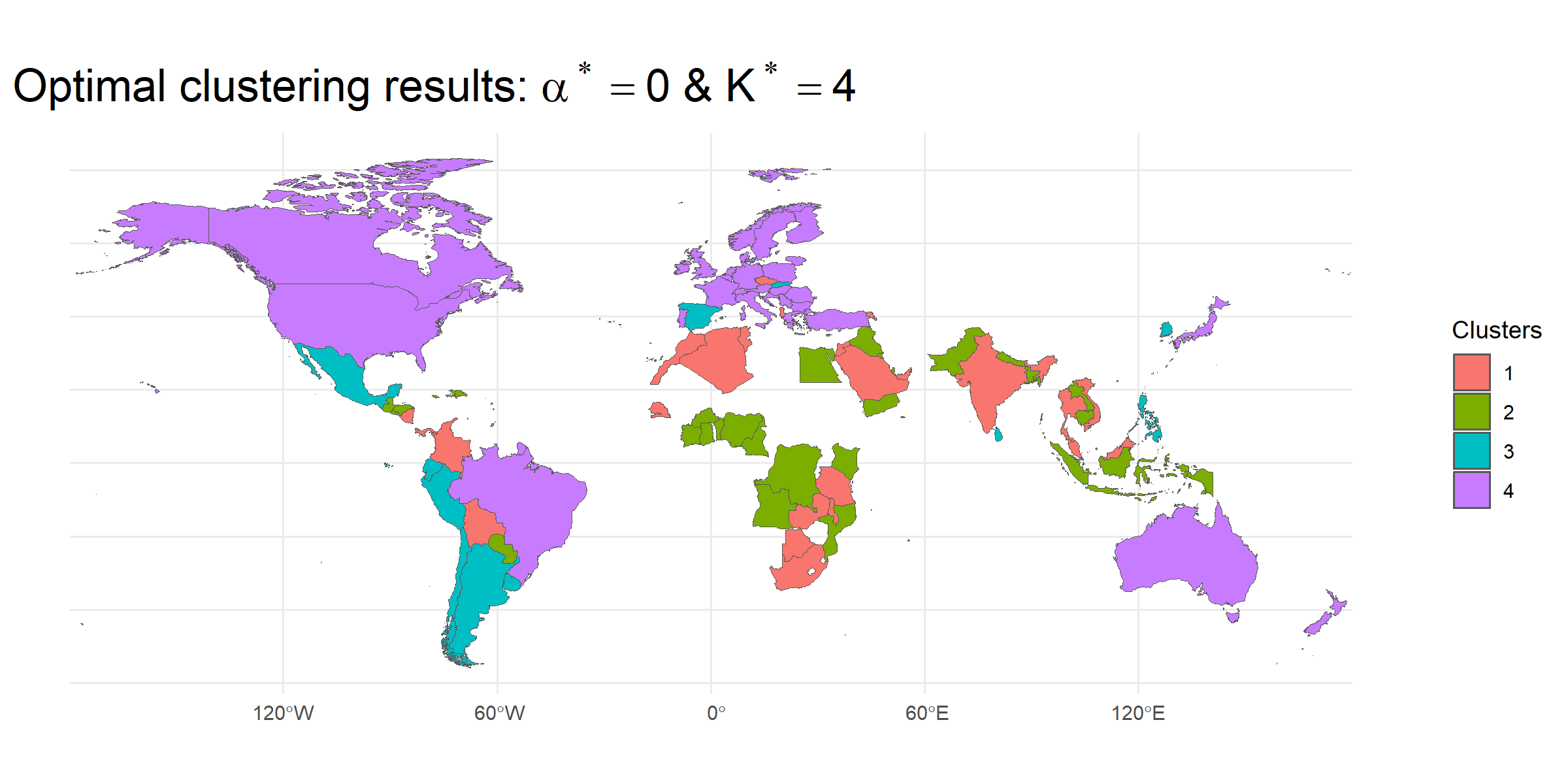}
	\caption{Map of clustering partitions obtained by setting $\alpha = 0$ and $K = 4$ and using as data the share of people in the 2022 survey with medium-high and high climate change awareness.}
	\label{figB:map_k4_a0}
\end{figure}

\begin{figure}[!htbp]
	\centering
	\includegraphics[width=1\columnwidth]{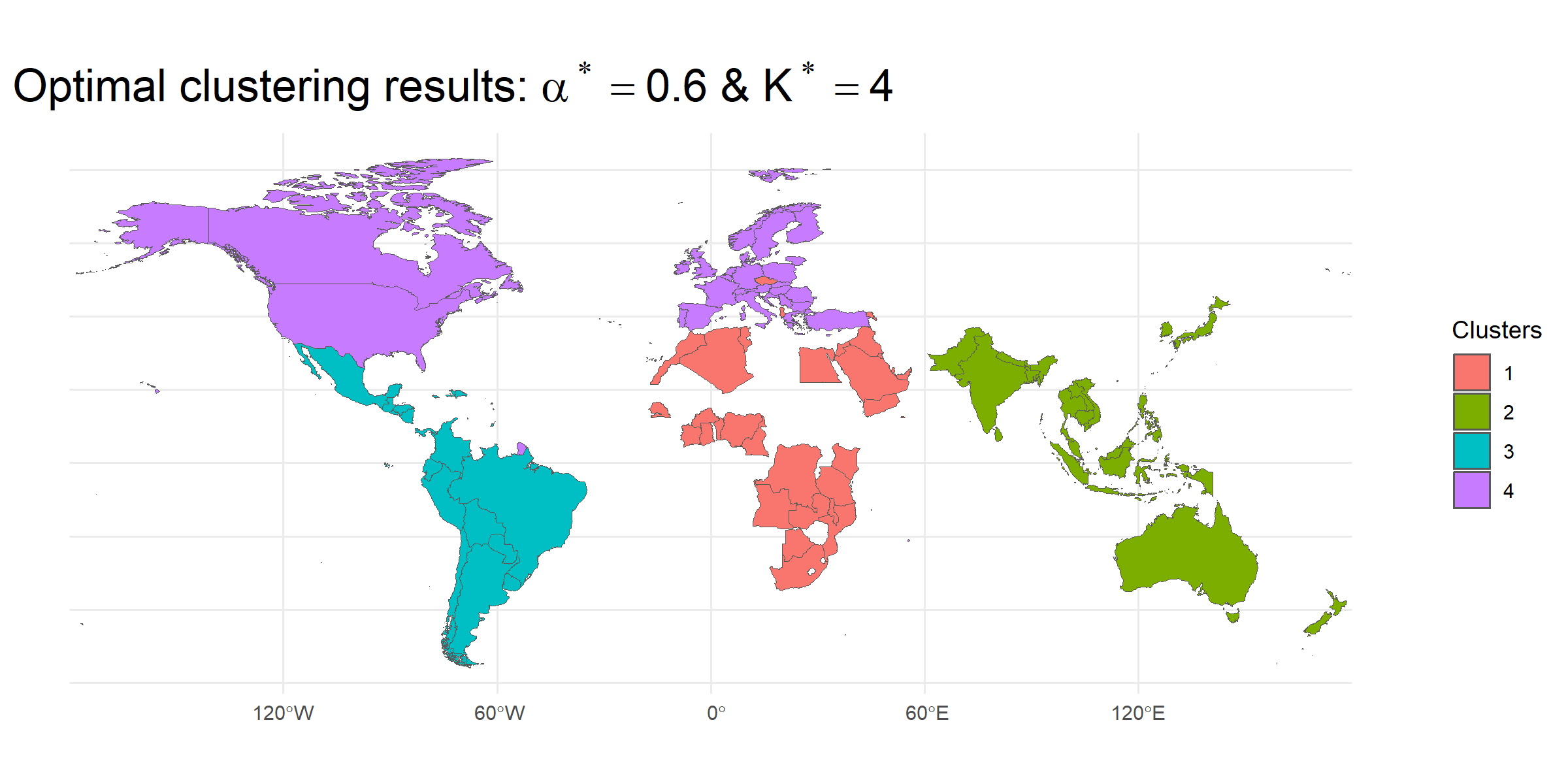}
	\caption{Map of clustering partitions obtained by setting $\alpha = 0.60$ and $K = 4$ and using as data the share of people in the 2022 survey with medium-high and high climate change awareness.}
	\label{figB:map_k4_a0.6}
\end{figure}

\begin{figure}[!htbp]
	\centering
	\includegraphics[width=1\columnwidth]{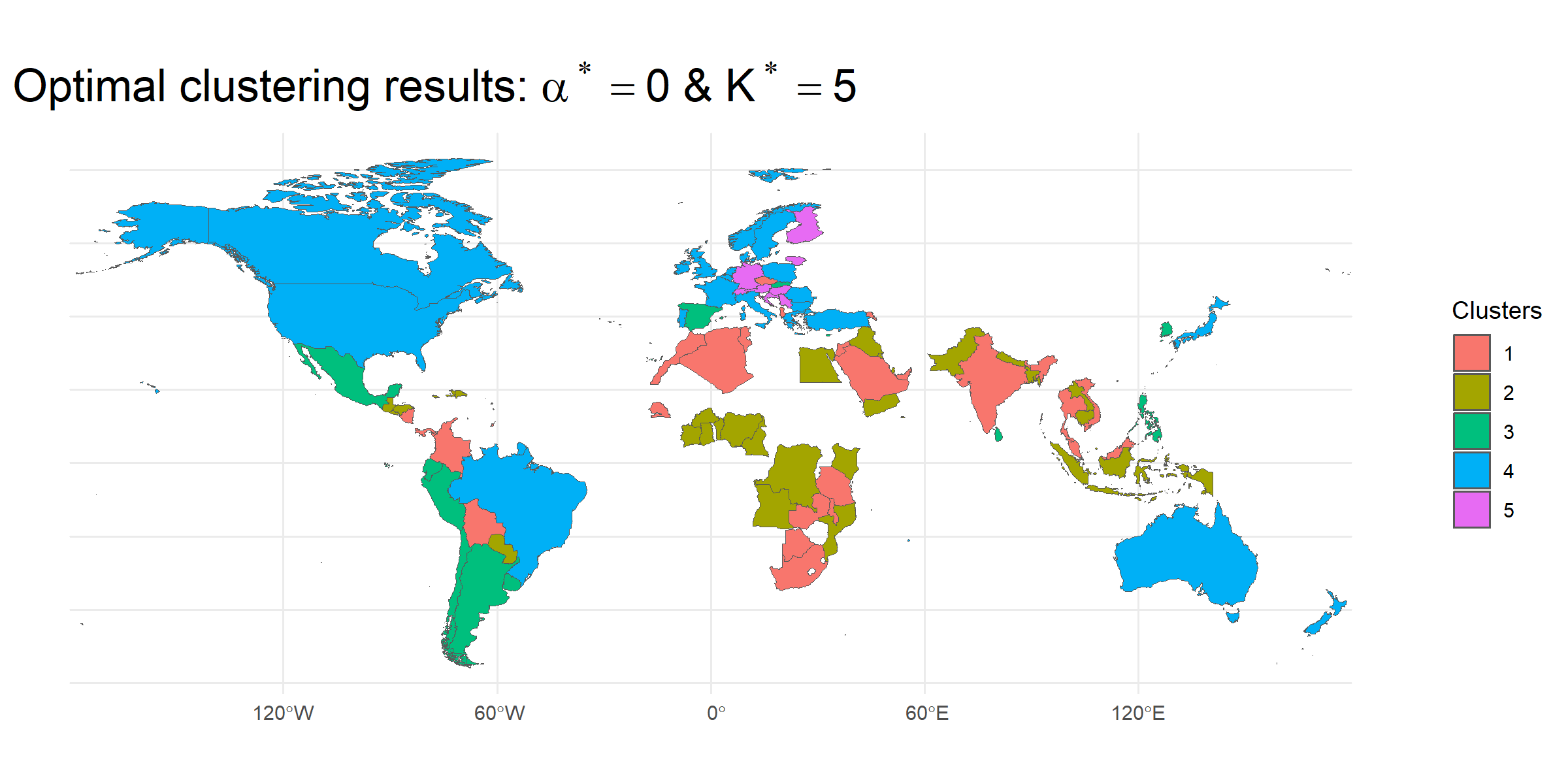}
	\caption{Map of clustering partitions obtained by setting $\alpha = 0$ and $K = 5$ and using as data the share of people in the 2022 survey with medium-high and high climate change awareness.}
	\label{figB:map_k5_a0}
\end{figure}

\begin{figure}[!htbp]
	\centering
	\includegraphics[width=1\columnwidth]{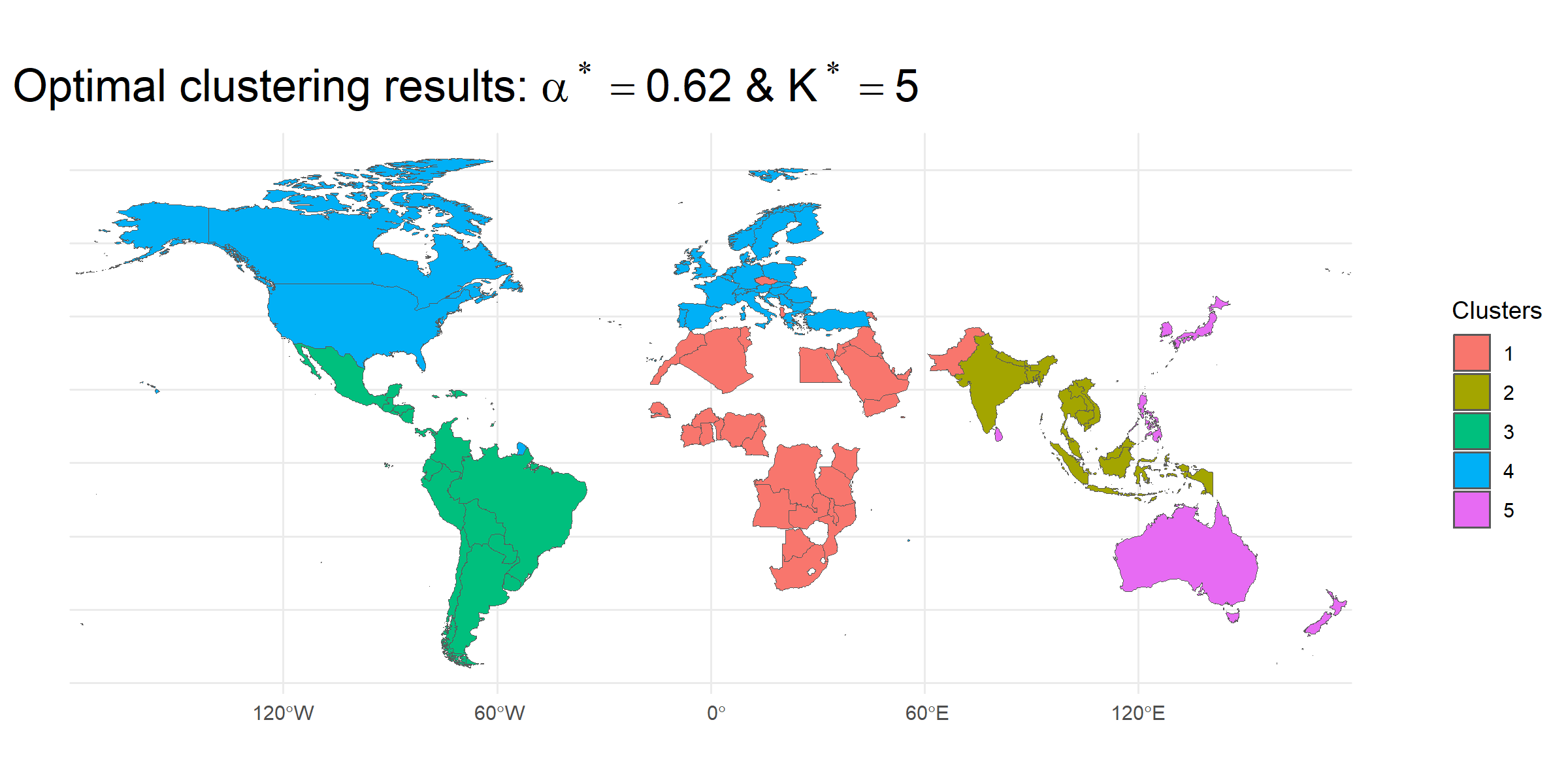}
	\caption{Map of clustering partitions obtained by setting $\alpha = 0.62$ and $K = 5$ and using as data the share of people in the 2022 survey with medium-high and high climate change awareness.}
	\label{figB:map_k5_a0.62}
\end{figure}

\clearpage
\section{Robustness analysis 2: spatial hierarchical clustering using low and medium-low climate change awareness + climate-related + socio-economic information}

\begin{figure}[!htbp]
	\centering
	\includegraphics[width=1\columnwidth]{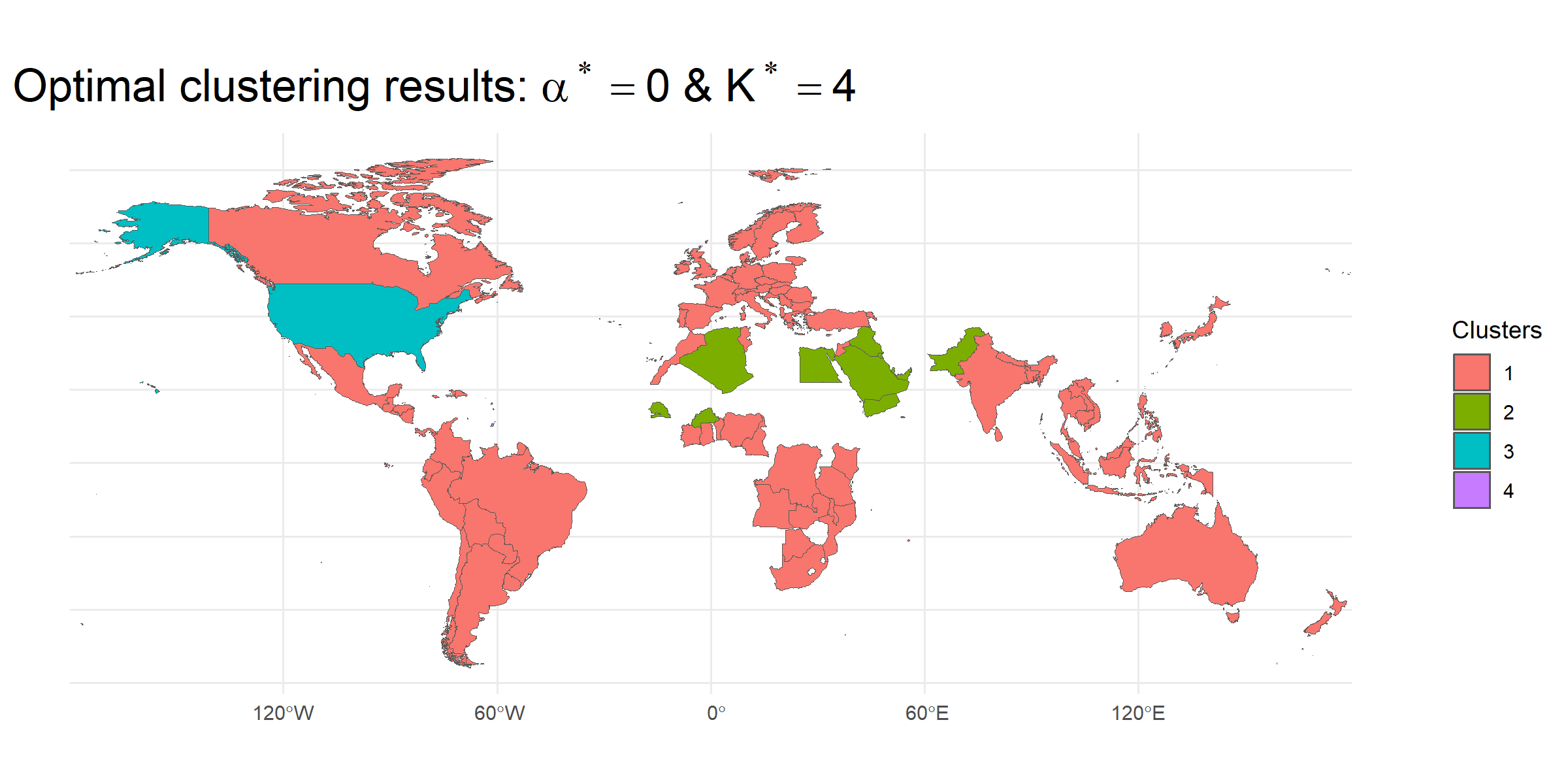}
	\caption{Map of clustering partitions obtained by setting $\alpha = 0$ and $K = 4$ and using information on the share of people in the 2022 survey with medium-low and low climate change awareness, on climate-related features and socio-economic features.}
	\label{figC:map_k4_a0}
\end{figure}

\begin{figure}[!htbp]
	\centering
	\includegraphics[width=1\columnwidth]{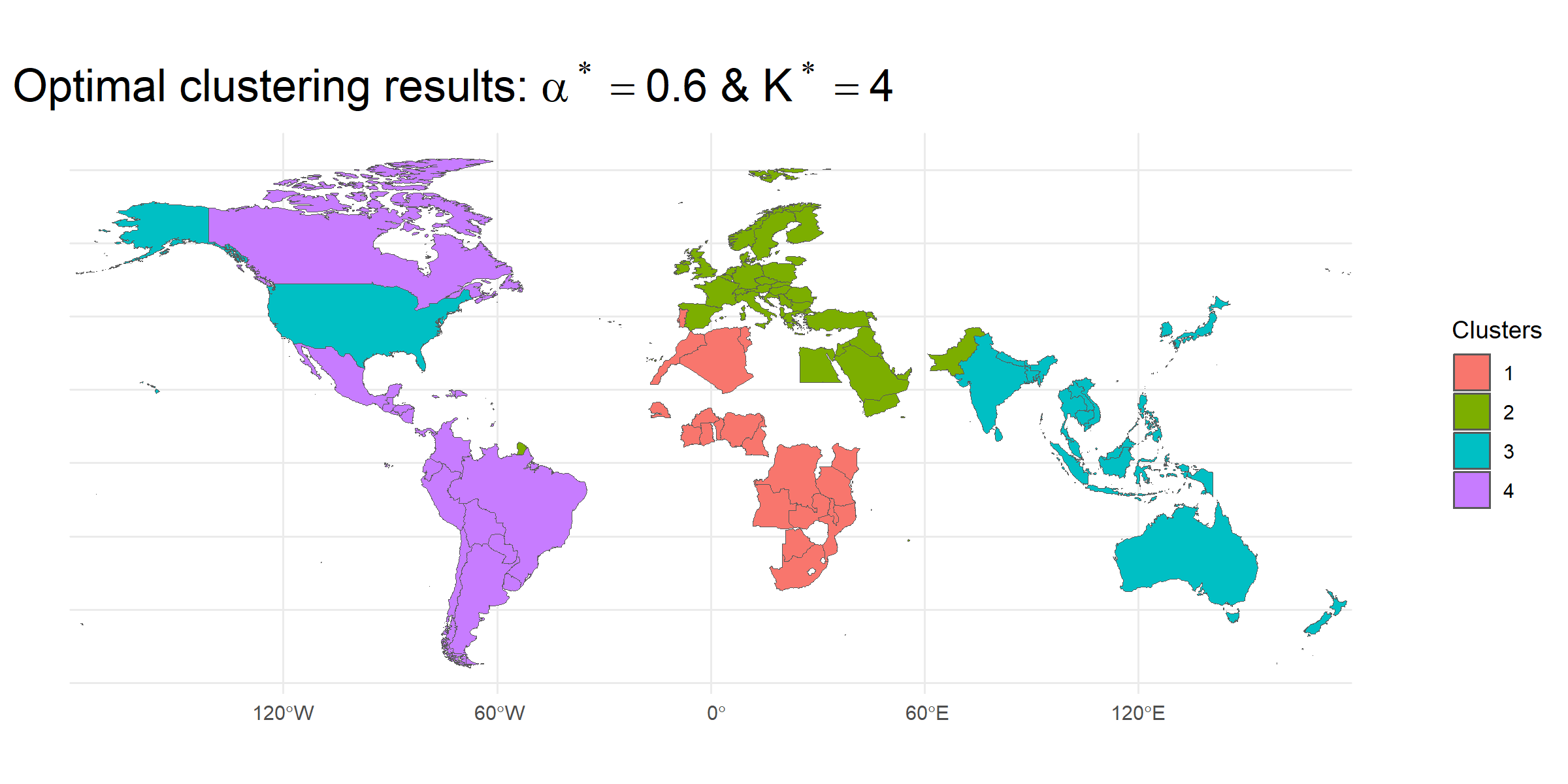}
	\caption{Map of clustering partitions obtained by setting $\alpha = 0.60$ and $K = 4$ and using information on the share of people in the 2022 survey with medium-low and low climate change awareness, on climate-related features and socio-economic features.}
	\label{figC:map_k4_a0.6}
\end{figure}

\begin{figure}[!htbp]
	\centering
	\includegraphics[width=1\columnwidth]{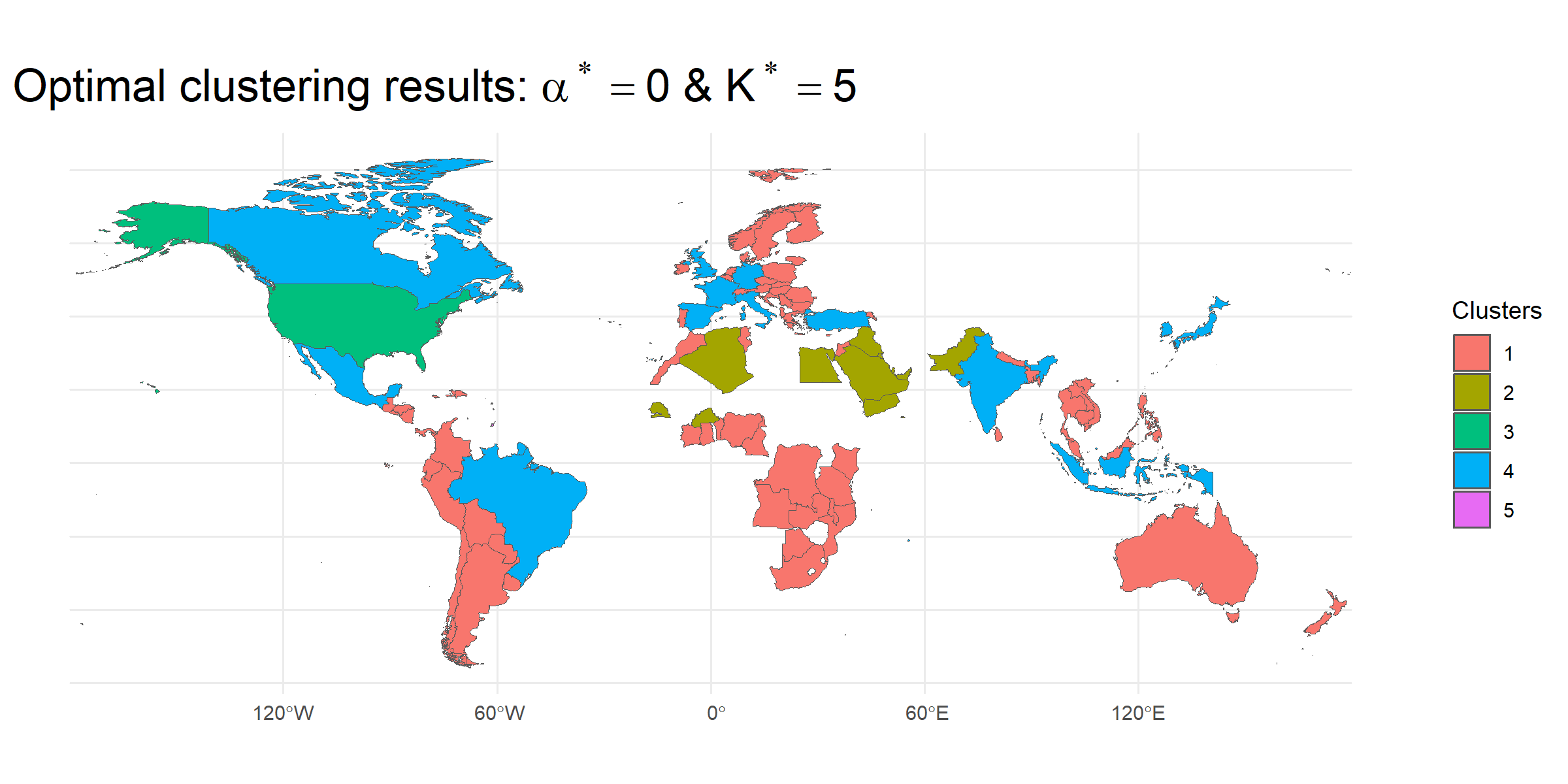}
	\caption{Map of clustering partitions obtained by setting $\alpha = 0$ and $K = 5$ and using information on the share of people in the 2022 survey with medium-low and low climate change awareness, on climate-related features and socio-economic features.}
	\label{figC:map_k5_a0}
\end{figure}

\begin{figure}[!htbp]
	\centering
	\includegraphics[width=1\columnwidth]{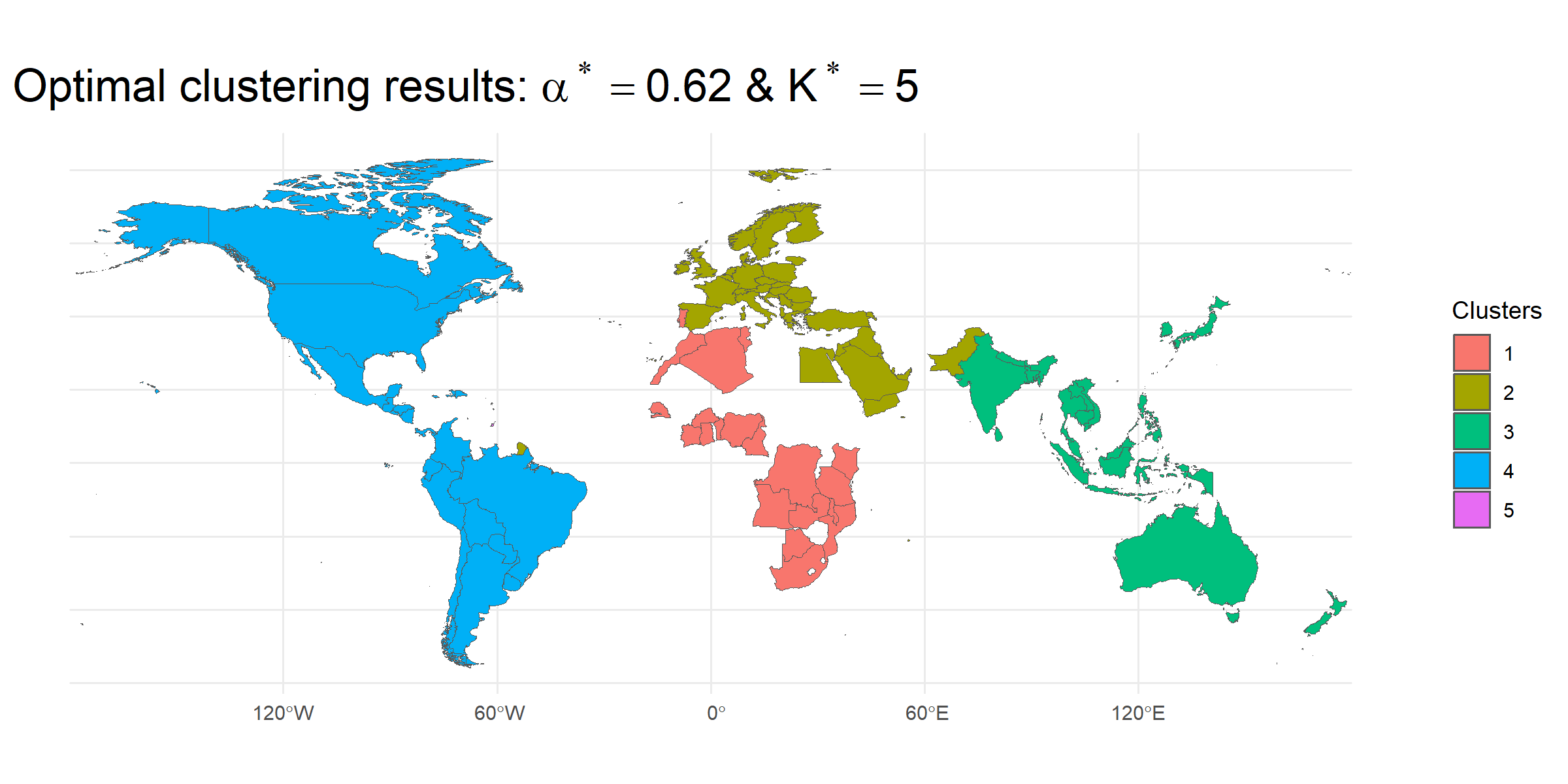}
	\caption{Map of clustering partitions obtained by setting $\alpha = 0.62$ and $K = 5$ and using information on the share of people in the 2022 survey with medium-low and low climate change awareness, on climate-related features and socio-economic features.}
	\label{figC:map_k5_a0.62}
\end{figure}

\clearpage
\section{Robustness analysis 2: spatial hierarchical clustering using low and medium-low climate change awareness + socio-economic information}

\begin{figure}[!htbp]
	\centering
	\includegraphics[width=1\columnwidth]{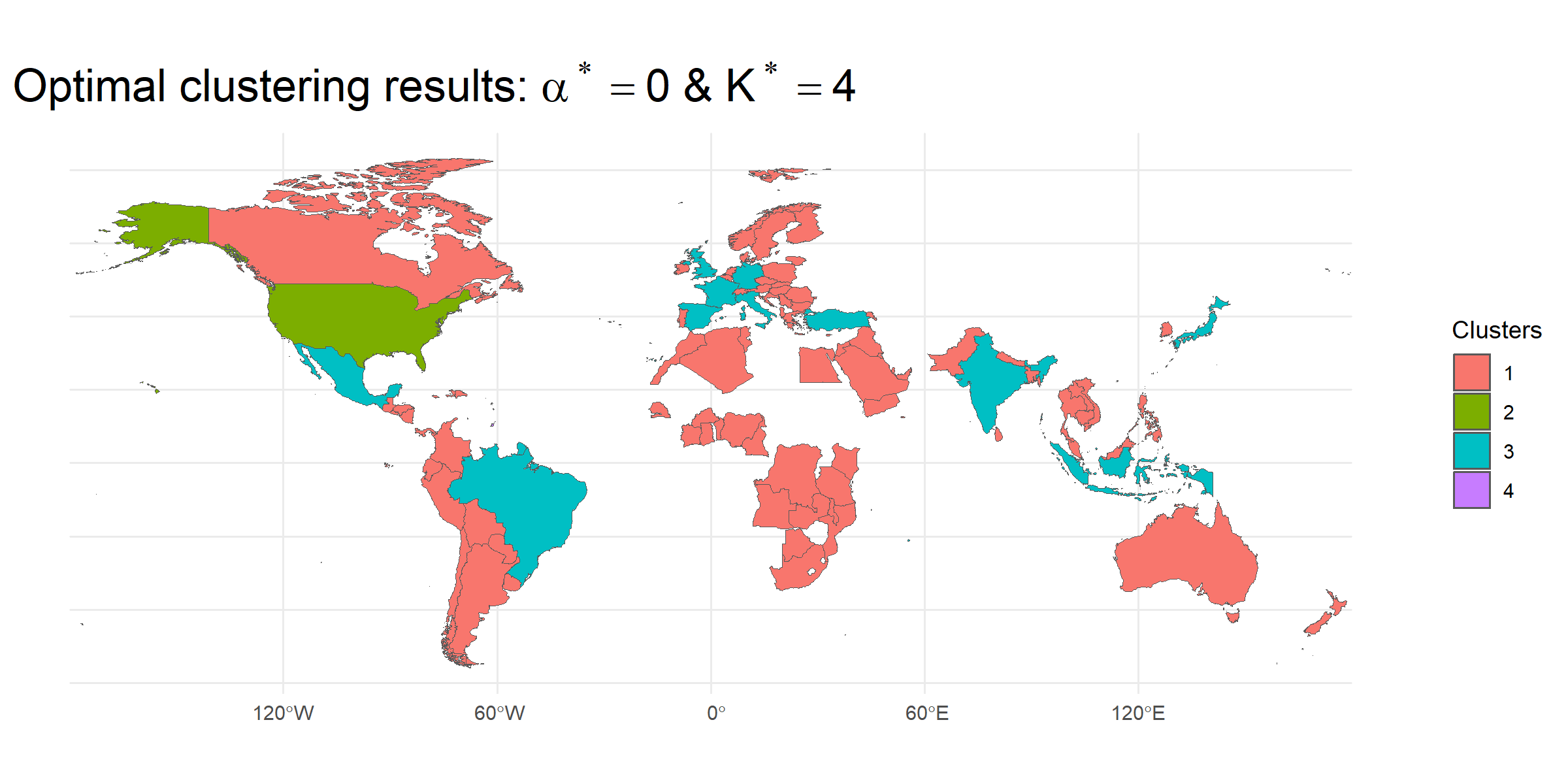}
	\caption{Map of clustering partitions obtained by setting $\alpha = 0$ and $K = 4$ and using information on the share of people in the 2022 survey with medium-low and low climate change awareness and socio-economic features.}
	\label{figD:map_k4_a0}
\end{figure}

\begin{figure}[!htbp]
	\centering
	\includegraphics[width=1\columnwidth]{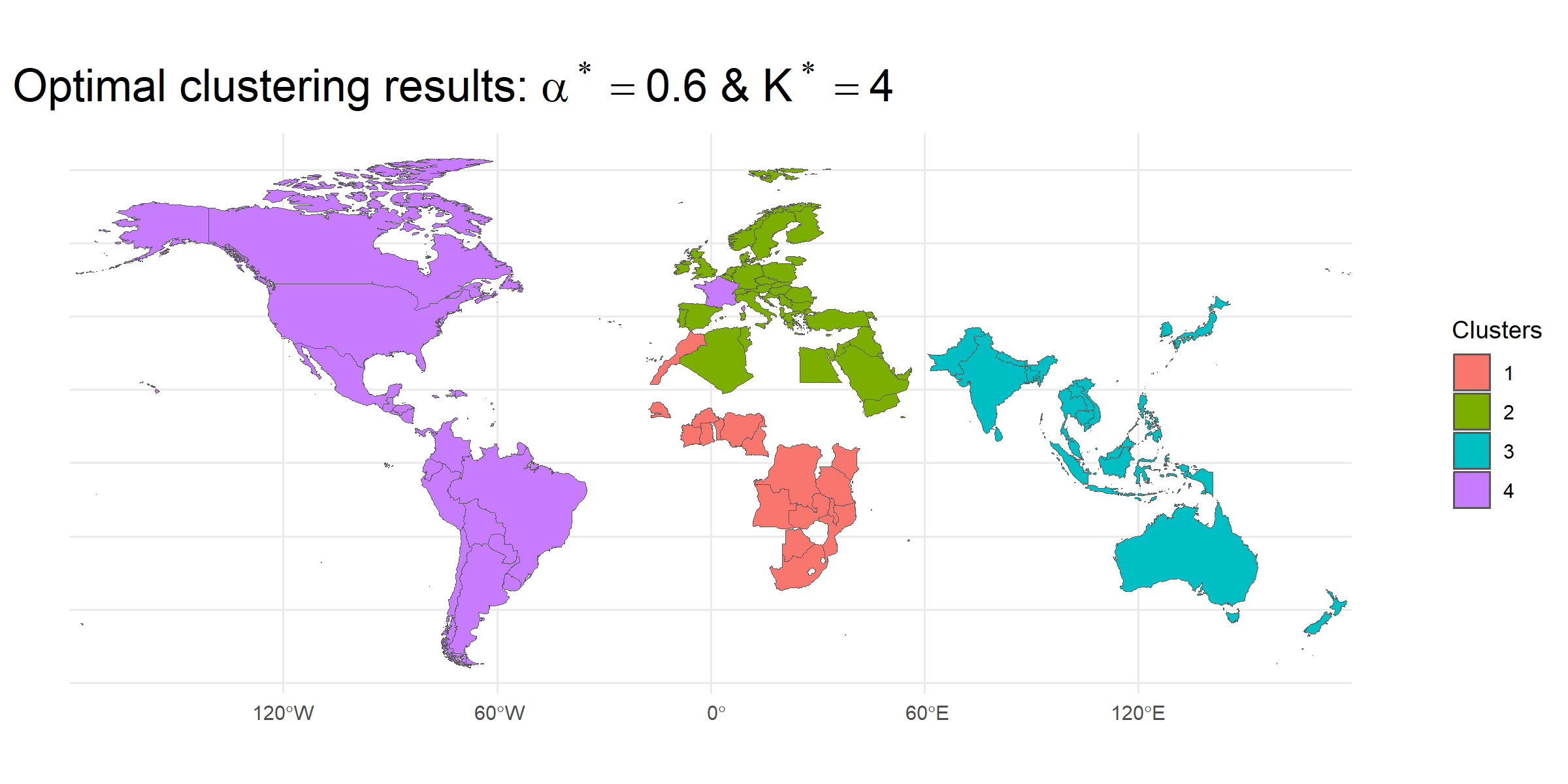}
	\caption{Map of clustering partitions obtained by setting $\alpha = 0.60$ and $K = 4$ and using information on the share of people in the 2022 survey with medium-low and low climate change awareness and socio-economic features.}
	\label{figD:map_k4_a0.6}
\end{figure}

\begin{figure}[!htbp]
	\centering
	\includegraphics[width=1\columnwidth]{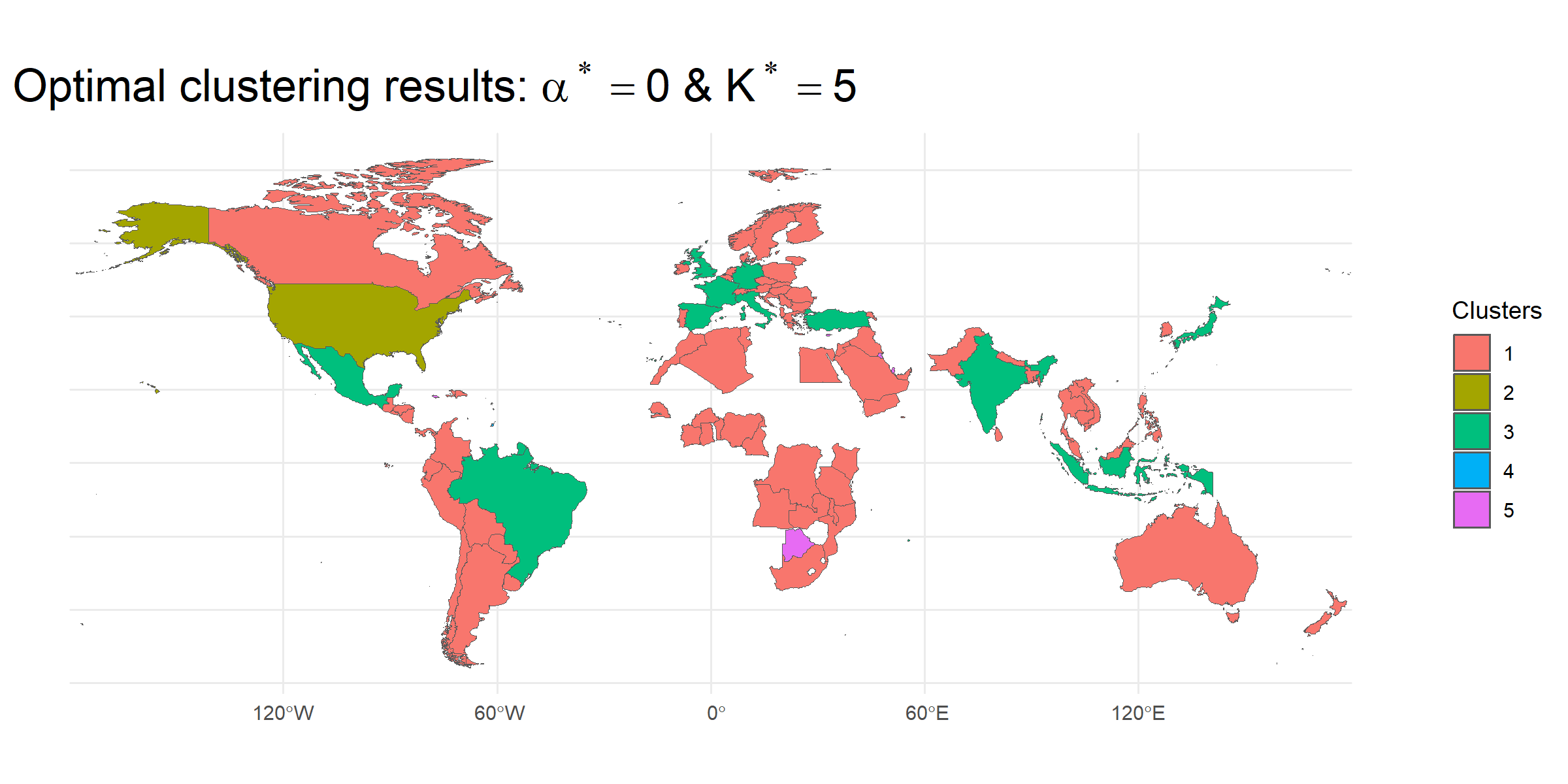}
	\caption{Map of clustering partitions obtained by setting $\alpha = 0$ and $K = 5$ and using information on the share of people in the 2022 survey with medium-low and low climate change awareness and socio-economic features.}
	\label{figD:map_k5_a0}
\end{figure}

\begin{figure}[!htbp]
	\centering
	\includegraphics[width=1\columnwidth]{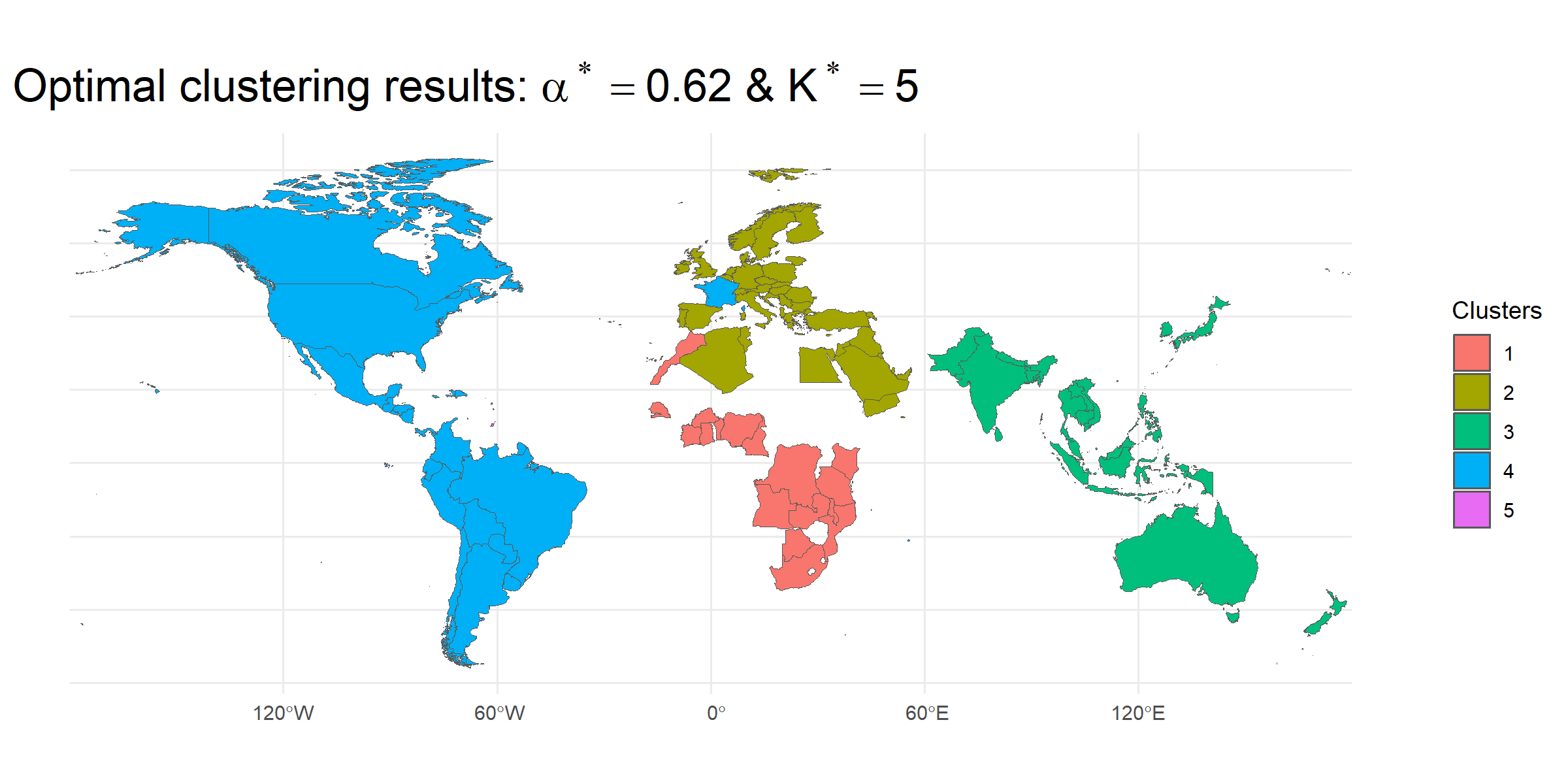}
	\caption{Map of clustering partitions obtained by setting $\alpha = 0.62$ and $K = 5$ and using information on the share of people in the 2022 survey with medium-low and low climate change awareness and socio-economic features.}
	\label{figD:map_k5_a0.62}
\end{figure}

\clearpage
\section{Robustness analysis 2: spatial hierarchical clustering using low and medium-low climate change awareness + climate-related information}

\begin{figure}[!htbp]
	\centering
	\includegraphics[width=1\columnwidth]{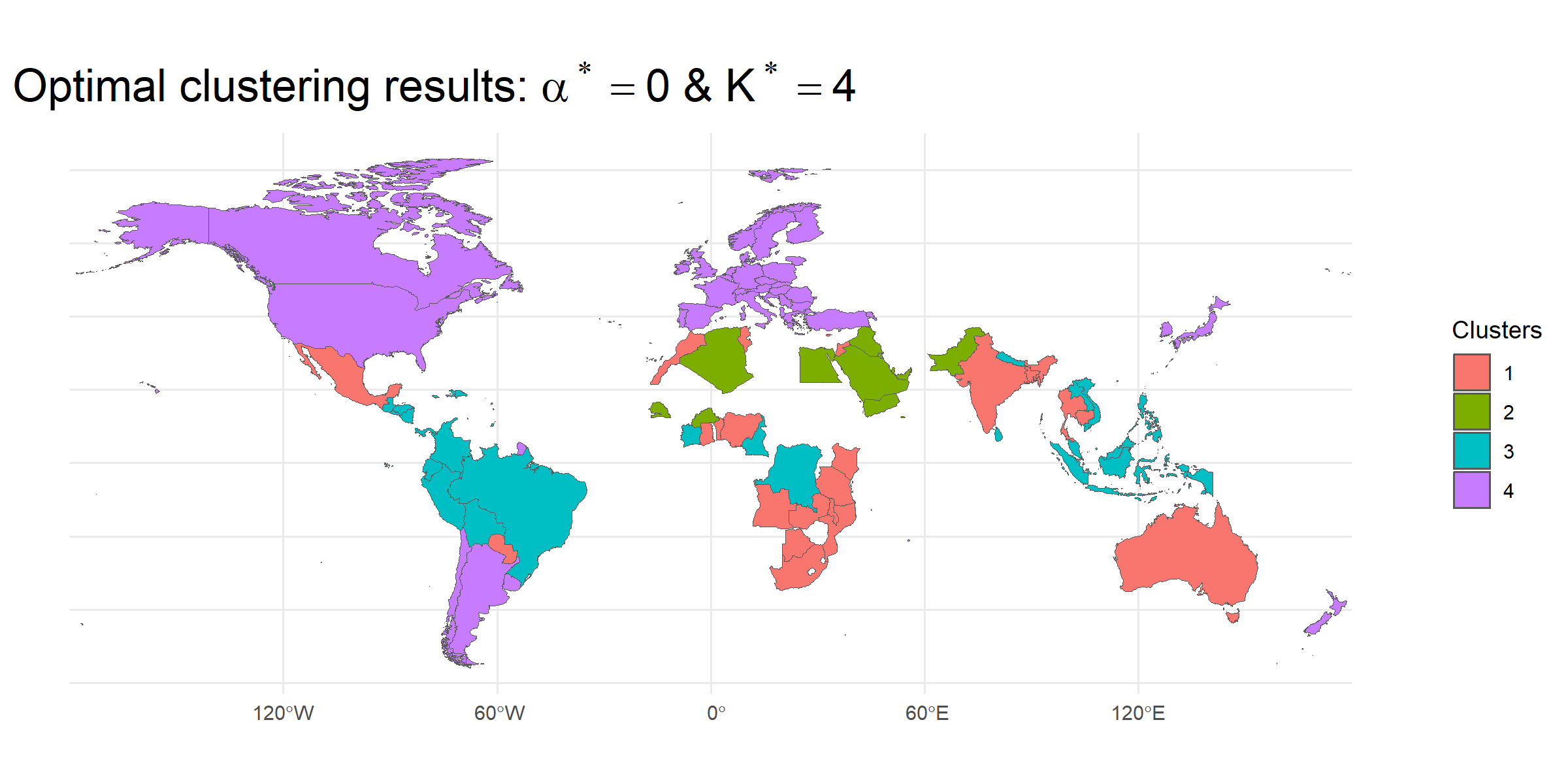}
	\caption{Map of clustering partitions obtained by setting $\alpha = 0$ and $K = 4$ and using information on the share of people in the 2022 survey with medium-low and low climate change awareness and climate-related features.}
	\label{figE:map_k4_a0}
\end{figure}

\begin{figure}[!htbp]
	\centering
	\includegraphics[width=1\columnwidth]{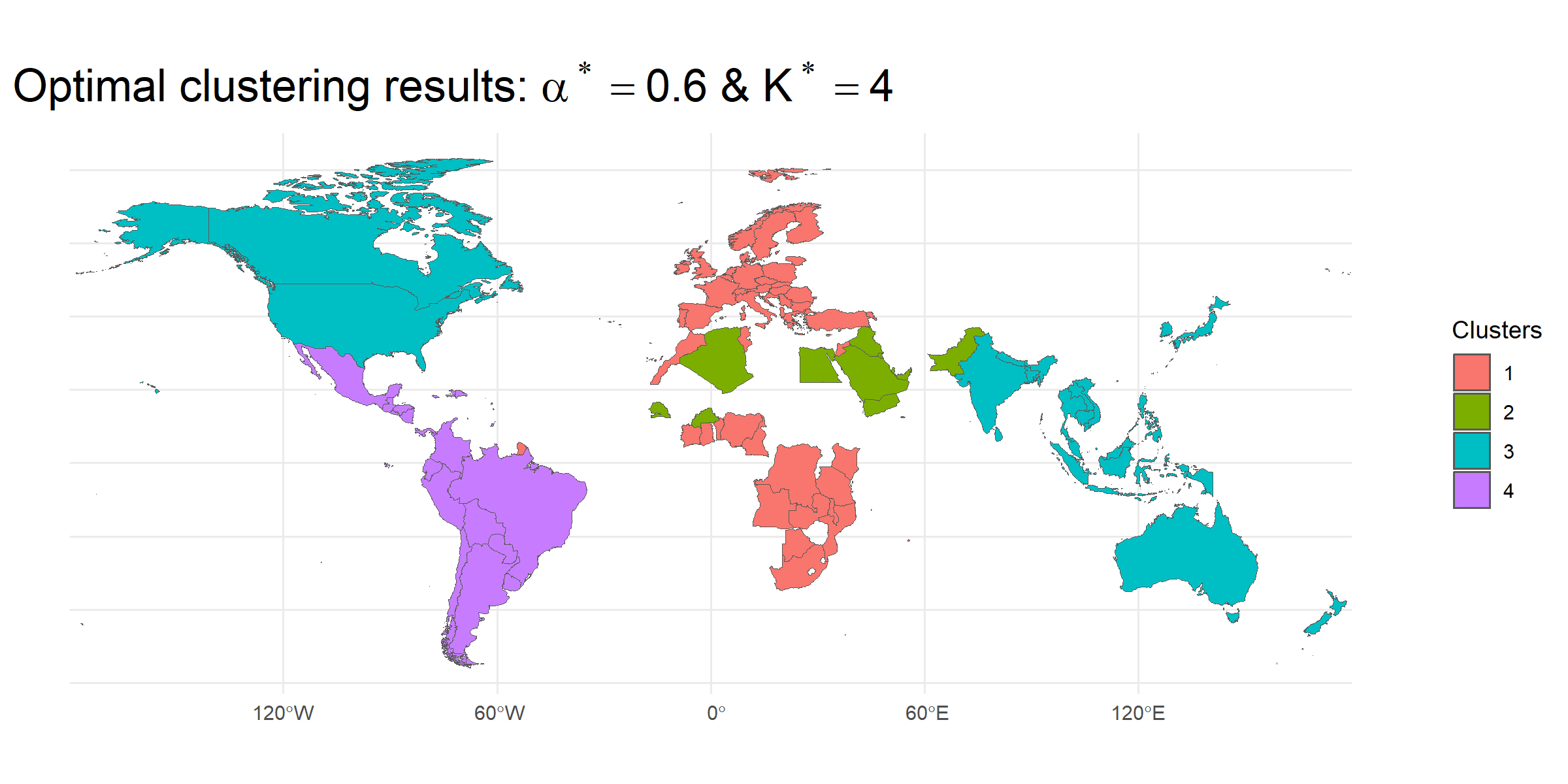}
	\caption{Map of clustering partitions obtained by setting $\alpha = 0.60$ and $K = 4$ and using information on the share of people in the 2022 survey with medium-low and low climate change awareness and climate-related features.}
	\label{figE:map_k4_a0.6}
\end{figure}

\begin{figure}[!htbp]
	\centering
	\includegraphics[width=1\columnwidth]{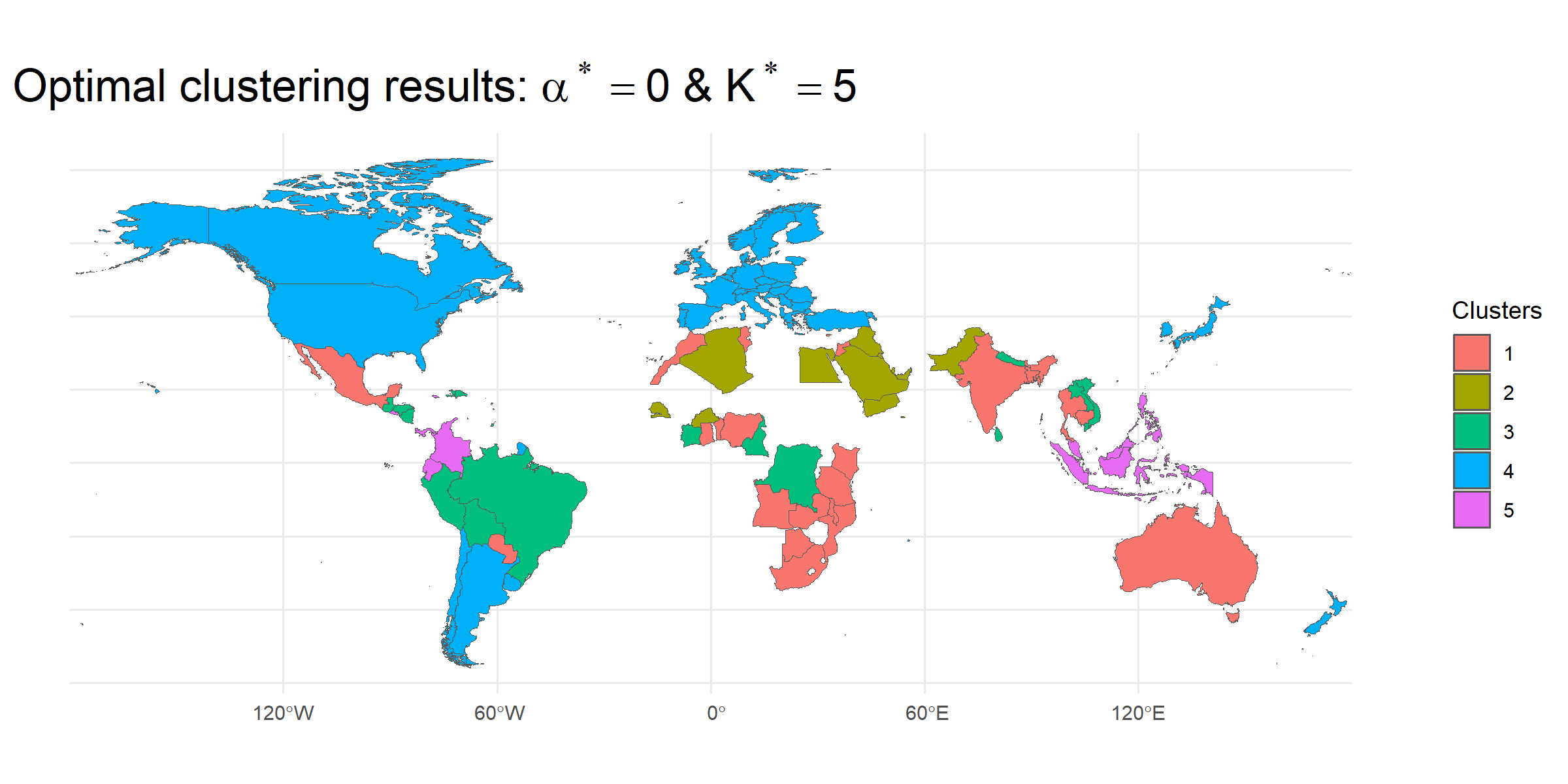}
	\caption{Map of clustering partitions obtained by setting $\alpha = 0$ and $K = 5$ and using information on the share of people in the 2022 survey with medium-low and low climate change awareness and climate-related features.}
	\label{figE:map_k5_a0}
\end{figure}

\begin{figure}[!htbp]
	\centering
	\includegraphics[width=1\columnwidth]{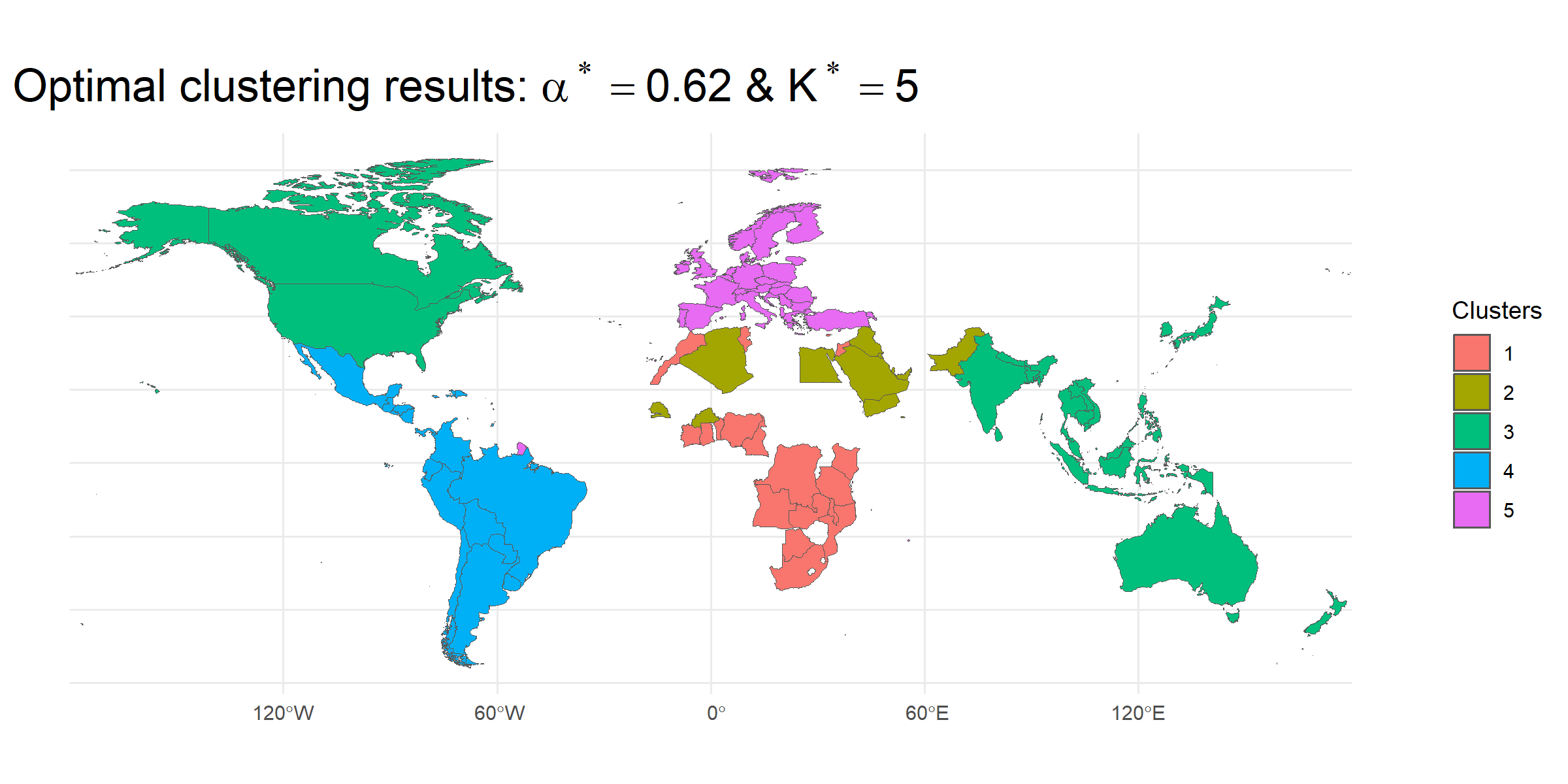}
	\caption{Map of clustering partitions obtained by setting $\alpha = 0.62$ and $K = 5$ and using information on the share of people in the 2022 survey with medium-low and low climate change awareness and climate-related features.}
	\label{figE:map_k5_a0.62}
\end{figure}

\end{document}